\documentclass[twocolumn]{aastex631}

\usepackage{amssymb, amsmath}
\usepackage{bm}
\usepackage{booktabs}
\usepackage[caption=false]{subfig}
\usepackage{float}

\newcommand{\bagpipes}{\textsc{BAGPIPES}}
\newcommand{\lya}{Ly$\alpha$}
\newcommand{\sqrchi}{$\chi^2$}
\newcommand{\virus}{VIRUS}
\newcommand{\nep}{NEP}
\newcommand{\hto}{H20}
\newcommand{\EW}{$W_{Ly\alpha}$}

\shortauthors{Ch\'{a}vez Ortiz et al.}

\begin{document}
 \pdfoutput=1
\title{Introducing the Texas Euclid Survey for Lyman Alpha (TESLA) Survey: Initial Study Correlating Galaxy Properties to Lyman-Alpha Emission}

\author[0000-0003-2332-5505]{\'{O}scar A. Ch\'{a}vez Ortiz}
\affiliation{Department of Astronomy, The University of Texas at Austin, 2515 Speedway Boulevard, Austin, TX 78712, USA}

\author[0000-0001-8519-1130]{Steven L. Finkelstein}
\affiliation{Department of Astronomy, The University of Texas at Austin, 2515 Speedway Boulevard, Austin, TX 78712, USA}

\author[0000-0002-8925-9769]{Dustin Davis}
\affiliation{Department of Astronomy, The University of Texas at Austin, 2515 Speedway Boulevard, Austin, TX 78712, USA}

\author[0000-0002-9393-6507]{Gene Leung}
\affiliation{Department of Astronomy, The University of Texas at Austin, 2515 Speedway Boulevard, Austin, TX 78712, USA}

\author[0000-0002-2307-0146]{Erin Mentuch Cooper}
\affiliation{Department of Astronomy, The University of Texas at Austin, 2515 Speedway Boulevard, Austin, TX 78712, USA}

\author[0000-0002-9921-9218]{Micaela Bagley}
\affiliation{Department of Astronomy, The University of Texas at Austin, 2515 Speedway Boulevard, Austin, TX 78712, USA}

\author[0000-0003-2366-8858]{Rebecca Larson}
\altaffiliation{NSF Graduate Fellow}
\affiliation{Department of Astronomy, The University of Texas at Austin, 2515 Speedway Boulevard, Austin, TX 78712, USA}

\author[0000-0002-0930-6466]{Caitlin M. Casey}
\affiliation{Department of Astronomy, The University of Texas at Austin, 2515 Speedway Boulevard, Austin, TX 78712, USA}

\author[0000-0002-3912-9368]{Adam P. McCarron}
\affiliation{Department of Astronomy, The University of Texas at Austin, 2515 Speedway Boulevard, Austin, TX 78712, USA}

\author[0000-0002-8433-8185]{Karl Gebhardt}
\affiliation{Department of Astronomy, The University of Texas at Austin, 2515 Speedway Boulevard, Austin, TX 78712, USA}

\author[0000-0002-4162-6523]{Yuchen Guo}
\affiliation{Department of Astronomy, The University of Texas at Austin, 2515 Speedway Boulevard, Austin, TX 78712, USA}

\author[0000-0001-5561-2010]{Chenxu Liu}
\affiliation{South-Western Institute for Astronomy Research, Yunnan University, Kunming, Yunnan, 650500, People’s Republic of China}
\affiliation{Department of Astronomy, The University of Texas at Austin, 2515 Speedway Boulevard, Austin, TX 78712, USA}

\author[0000-0003-4323-0597]{Isaac Laseter}
\affiliation{Department of Astronomy, University of Wisconsin-Madison, 475 North Charter Street, Madison, WI 53706, USA}

\author{Jason Rhodes}
\affiliation{Jet Propulsion Laboratory, California Institute of Technology, 4800 Oak Grove Dr, Pasadena, CA 91109
}
\author{Ralf Bender}
\affiliation{Max-Planck-Institut f\"ur extraterrestrische Physik,
Postfach 1312, Giessenbachstr., 85748 Garching, Germany}

\author{Max Fabricius}
\affiliation{Max-Planck-Institut f\"ur extraterrestrische Physik,
Postfach 1312, Giessenbachstr., 85748 Garching, Germany}

\author[0000-0003-1198-831X]{Ariel G. S\'anchez}
\affiliation{Max-Planck-Institut f\"ur extraterrestrische Physik,
Postfach 1312, Giessenbachstr., 85748 Garching, Germany}

\author{Claudia Scarlata}
\affiliation{Minnesota Institute for Astrophysics, University of Minnesota, 116 Church Street SE, Minneapolis, MN 55455, United States}

\author{Peter Capak}
\affiliation{Infrared Processing and Analysis Center, 1200 E California Blvd, Pasadena, CA 91125}

\author[0000-0001-5680-2326]{Lukas Zalesky}
\affiliation{University of Hawaii Institute for Astronomy, 2680 Woodlawn Dr, Honolulu, HI 96822}

\author{David Sanders}
\affiliation{University of Hawaii Institute for Astronomy, 2680 Woodlawn Dr, Honolulu, HI 96822}

\author[0000-0003-2274-0301]{Istvan Szapudi}
\affiliation{University of Hawaii Institute for Astronomy, 2680 Woodlawn Dr, Honolulu, HI 96822}

\author{Eric Baxter}
\affiliation{University of Hawaii Institute for Astronomy, 2680 Woodlawn Dr, Honolulu, HI 96822}

\author[0000-0003-0639-025X]{Conor McPartland}
\affiliation{Cosmic Dawn Center (DAWN), Niels Bohr Institute, University of Copenhagen, Jagtvej 128, København N, DK-2200, Denmark}

\author[0000-0003-1614-196X]{John R. Weaver}
\affiliation{Department of Astronomy, University of Massachusetts, Amherst, MA 01003, USA}
\affiliation{Cosmic Dawn Center (DAWN), Copenhagen, Denmark}
\affiliation{Niels Bohr Institute, University of Copenhagen, Jagtvej 128, København N, DK-2200, Denmark}

\author[0000-0003-3631-7176]{Sune Toft}
\affiliation{Cosmic Dawn Center (DAWN), Copenhagen, Denmark}
\affiliation{Niels Bohr Institute, University of Copenhagen, Jagtvej 128, København N, DK-2200, Denmark}
\author{Bahram Mobasher}

\author{Nao Suzuki}
\affiliation{E.O. Lawrence Berkeley National Lab, Berkeley, CA 94720, USA}

\author[0000-0003-3691-937X]{Nima Chartab}
\affiliation{The Observatories of the Carnegie Institution for Science, 813 Santa Barbara St., Pasadena, CA 91101, USA}

\begin{abstract}
We present the Texas Euclid Survey for Lyman-Alpha (TESLA), a spectroscopic survey in the 10 $deg^2$ of the \textit{Euclid} North Ecliptic Pole (NEP) field. Using TESLA, we study how the physical properties of Lyman-$\alpha$ emitters (LAEs) correlate with \lya\ emission to understand the escape of Ly$\alpha$ from galaxies at redshifts 2 -- 3.5. We present an analysis of 43 LAEs performed in the NEP field using early data from the TESLA survey. We use Subaru Hyper Suprime-Cam imaging in the $grizy$-bands, \textit{Spitzer}/IRAC channels 1 and 2 from the Hawaii 20-$deg^2$ (H20) survey and spectra acquired by the Visible Integral-Field Replicable Unit Spectrograph (VIRUS) on the Hobby-Eberly Telescope. We perform spectral energy distribution (SED) fitting to compute the galaxy properties of 43 LAEs, and study correlations between stellar mass, star formation rate (SFR), and dust, to the \lya\ rest-frame equivalent width (\EW). We uncover marginal (1$\sigma$ significance) correlations between stellar mass and \EW, and star formation rate (SFR) and \EW, with a Spearman correlation coefficient of -0.$34_{-.14}^{+.17}$ and -0.$37_{-.14}^{+.16}$ respectively. We show that the \EW\ distribution of the 43 LAEs is consistent with being drawn from an exponential distribution with an e-folding scale of $W_0 = 150$ \AA. Once complete the TESLA survey will enable the study of $\gtrsim$ 50,000 LAEs to more explore correlations between galaxy properties and \EW. The large sample size will allow the construction of a predictive model for the \EW\ as a function of SED-derived galaxy properties, which could be used to improve Ly$\alpha$-based constraints on reionization.

\end{abstract}


\vspace*{20mm}
\section{Introduction} \label{sec:intro}
The Universe's dark ages ended when the first stars and galaxies start emitting radiation. Eventually, the energy from the emitted photons becomes large enough to ionize the intergalactic medium (IGM). The time the IGM transitioned from neutral to ionized is known as the Epoch of Reionization. Evidence from the Gunn-Peterson trough of the residual neutral fraction from quasar spectra point to reionization ending around z $\approx$ 5.5, but the precise time of its onset, duration and underlying driving mechanism are still unsolved questions \citep[e.g.][]{gunn_peterson, QSO_reionization, QSO_ref2}. Measuring the duration of reionization allows us to constrain the sources that contributed to ionizing the IGM. At one point it was thought that active galactic nuclei (AGN) were responsible for reionization since they are very luminous and produce a vast amount of ionizing photons, but the number density of AGN at very high redshift is not sufficient to cause reionization to end by z $\approx$ 5.5 \citep[e.g.,][]{Not_AGN_Ion1, Not_AGN_ref2}. Evidence points to star-forming galaxies being the main contributors of the ionizing photons that reionized the universe, but we still do not know whether bright or faint galaxies are responsible \citep[][]{Not_AGN_ref2, Naidu2020, Robertson2015}. To answer which galaxies are responsible for driving reionization we need to make robust measurements of the neutral hydrogen fraction during the epoch of reionization since the competing models predict different reionization histories \citep[e.g.][]{Neutral_Fraction1, Neutral_Fraction2, Neutral_Fraction3, Not_AGN_ref2, Whitler2020, Hoag_2019}.

While many methods exist for measuring the neutral fraction during the epoch of reionization, such as using \ion{H}{1} 21cm intensity mapping, Gunn-Peterson trough measurements, Thompson Scattering of CMB photons \citep[][and references therein]{Lya_Review}, a sensitive method that can be applied now is using the \lya\ emission line to compute the neutral hydrogen fraction. \lya\ has been shown to be a very promising tracer for measuring the neutral hydrogen fraction as \lya\ should be fairly bright from the numerous star-forming galaxies present during the epoch of reionization \citep[e.g.][]{Bouwens, Finkelstein_2015}. \lya\ has the advantageous feature of being a resonant line, meaning that it is easily absorbed by neutral hydrogen and re-emitted in a random direction, significantly reducing its observability. We expect this scattering to occur frequently when in the presence of a neutral IGM since most of the hydrogen will be in the ground state. All these properties make \lya\ a reasonable tracer to compute the neutral hydrogen fraction as \lya\ can get absorbed easily by any neutral hydrogen in the surrounding area near where it was emitted \citep[e.g.][and references therein]{ME_1988, MR_2004, Reionization_Galaxies}. 

To measure the hydrogen neutral fraction using \lya\ emission requires us to know two things; the observed flux of \lya\ of a galaxy and the \lya\ flux post-ISM but pre-IGM of a galaxy. One can measure the observed flux of \lya\ using ground-based and space-based telescopes but there is no good method to quantify the escaped \lya\ strength of distant galaxies. This is due to internal processes within a galaxy such as dust content, ISM kinematics, and ISM geometry affecting the \lya\ flux that ultimately escapes from a galaxy \citep[e.g.][and references therein]{Wofford_2013, Rivera_2015, Trainor_2015, Reionization_Galaxies}. Since internal processes regulate the escape of \lya\ we cannot rely on traditional luminosity or SFR estimators to estimate the \lya\ flux of a galaxy. Thus, accurate measurements of the amount of \lya\ photons that escape a galaxy, to date, has been a limiting factor in previous studies \citep[e.g.][]{Pent_2018, Neutral_Fraction2, Jung_2020}, which typically use the measured \EW\ distribution at z $\approx$ 6 as an estimate of the \lya\ emission at z $> 6$. However, this method makes the assumption that the z $\approx$ 6 sample is able to represent the entire galaxy population at higher redshifts, which is not the case as small sample statistics and selection bias can influence which galaxies are selected and studied. Studying \lya\ emission at such high redshifts requires the selection of sources for spectroscopic follow-up, which imparts a selection bias towards the most luminous and massive sources. This is complicated further by different studies using different selection criteria when selecting candidates for spectroscopic follow-up (ie: color-cuts: \cite{Redshift2_6}, Narrow Band (NB) Imaging: \cite{Xshooter}, \cite{Lager_NB}). A complimentary, more comprehensive, and unbiased way of studying LAEs is needed to overcome these observational barriers.

We aim to overcome these observational barriers by using the Texas Euclid Search for Lyman-alpha (TESLA) Survey. When complete, TESLA will have data for $\sim$ 50,000 LAEs in the 10 $deg^2$ \textit{Euclid} North Ecliptic Pole (NEP) field centered on $\alpha(2000)$ = 270.0$^{\circ}$, $\delta(2000)$ = 66.6$^{\circ}$. TESLA's science mission is to study LAEs between $z =$ 1.9 -- 3.5, and determine how SED-derived galaxy properties correlate to the escaped \lya\ emission. We do this by pairing spectroscopic data provided by \virus\ on the Hobby-Eberly Telescope (HET) with deep Subaru Hyper Suprime-Cam (HSC) photometry and imaging from the Hawaii 20 $deg^2$ survey (\hto, \citet[][in prep]{Mcpartland_2022}). Our end goal is to make a predictive distribution of the \lya\ flux so that for any given set of SED-derived properties we can probabilistically determine the emerged \lya\ flux of a galaxy during the epoch of reionization. The TESLA survey is optimal for this analysis as galaxies between redshifts 1.9 -- 3.5 are in a fully ionized IGM at these redshifts. Another advantage of the TESLA survey is that previous LAE studies were limited by small sample sizes ($\sim$100-1000 LAEs) and inconsistent selection criteria between studies \citep[e.g.][]{Xshooter, LAEs_2_3, LAEs_21, LAEs_31}. The TESLA survey, once complete, will have roughly 10-100x the amount of LAEs than previous studies, all selected using the same methodology. The expanded sample size will allow us to more robustly explore \lya\ emission and how a given galaxy property hinders or promotes \lya\ escape. We will also probe any redshift evolution, if any, on the strength of \lya\ between redshifts 1.9 -- 3.5. This makes the TESLA survey the optimal survey to correlate \lya\ strength to SED-derived galaxy properties.

In this work, we introduce the TESLA survey and perform a study on a part of the NEP field where spectroscopic and imaging data were fully complete. We develop a framework that can find LAEs, acquire SED-derived galaxy properties, and explore initial correlations between galaxy properties and \lya\ rest-frame equivalent widths (\EW). The outline of the paper is as followed: \S \ref{sec:data} covers the data sets used in the TESLA survey. \S\ref{sec:methods} covers the methodology used in the analysis. \S\ref{sec:results} covers the results of the study. \S \ref{sec:discussion} covers a discussion of the results and \S\ref{sec:summary} summarizes our results and what we expect to see moving forward in the analysis with an expanded data set. We assume a cosmological model with $H_0$ = 70 km $s^{-1}$ $Mpc^{-1}$, $\Omega_{m, 0}$ = 0.3, and
$\Omega_{\Lambda, 0}$ = 0.7 and all magnitudes reported are expressed in the AB magnitude system. We define a galaxy to be an LAE if the galaxy has a spectroscopic redshift between 1.9 -- 3.5, to have a rest frame \lya\ equivalent width $> 20$ \AA, following \citet[e.g.,][]{rhoads00,Shapley_2003, Gawiser_2007}, and for the galaxy to not have any identifiable Active Galactic Nuclei (AGN) features, such as broad-line features.

\begin{figure*}
\includegraphics[width = .95\textwidth]{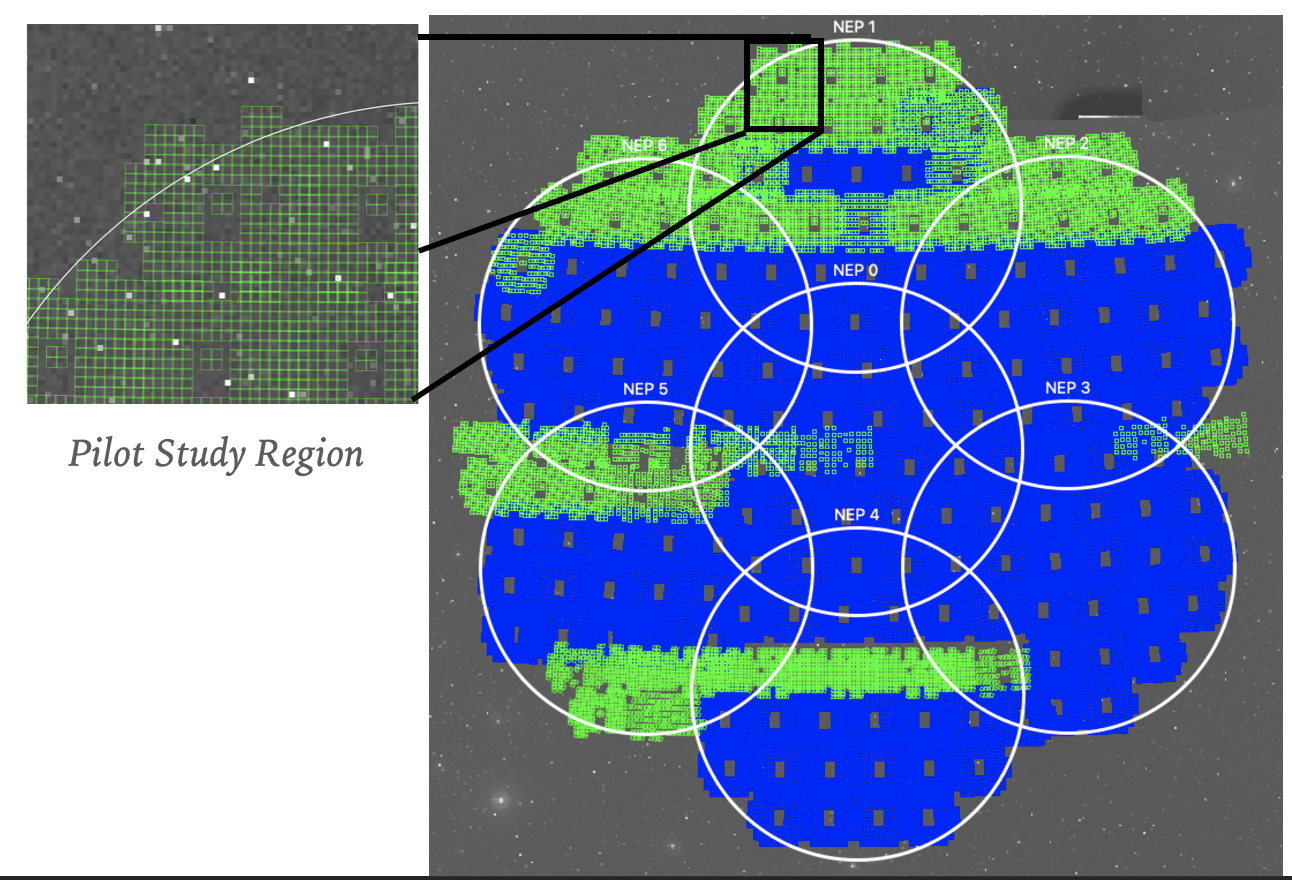}
\caption{An imaging footprint of the 10 $deg^2$ NEP field, which will also be deeply imaged by \textit{Euclid}. The green boxes are observed TESLA pointings, each pointing representing 4 shots, with each shot having a maximum of 78 IFUs covering 51$^{\prime\prime}$ x 51$^{\prime\prime}$. The green boxes show data that was collected between July 2018 to May 2020, and the blue boxes are data that will be collected in the coming years. The data used in this study can be seen in the zoom-in on the top left of the image and covers a 194 arcmin$^2$ area, with a total of 536 IFUs used for these observations. The white circles outline the H20 imaging survey fields. The full TESLA dataset, once complete, would uncover $\sim$ 50,000 LAEs. When combined with the H20 imaging, the TESLA survey will enable a robust study of LAEs galaxy properties to \lya\ emission strength between $z =$ 2--3.5.}
\label{fig:NEP_Footprint}
\end{figure*}

\section{Data} \label{sec:data}

\subsection{Introducing TESLA}

TESLA is an unbiased spectroscopic survey of the 10 $deg^2$ \textit{Euclid} North Ecliptic Pole (NEP) field using the Visible Integral-Field Replicable Unit Spectrograph (VIRUS) instrument on the Hobby-Eberly Telescope \cite{Hill_2021}. The photometric data for TESLA comes from the Hawaii-20 $deg^2$ (\hto) survey which is, at the time of this paper, acquiring 
deep optical photometry and imaging in the NEP field. The spectroscopic data provides us with the \lya\ flux and the galaxy's redshift. The photometric information is used to corroborate the emission line detected by the VIRUS instrument and to perform SED fitting to acquire a galaxy's global properties. The science goal of TESLA is to study the characteristics of roughly 50,000 LAEs between redshifts 1.9 -- 3.5 so that we can study the emerged \lya\ flux of LAEs without any significant IGM attenuation on the \lya\ emission. We aim to use any correlations between SED-derived galaxy properties to the \lya\ emission and generate a predictive distribution of \lya\ strength that is tied to global galaxy properties. 

The footprint of the \nep\ field can be seen in Figure~ \ref{fig:NEP_Footprint} where we show the TESLA VIRUS pointings, where individual IFUs are represented by the small square-like features in Figure~\ref{fig:NEP_Footprint}. The green pointings in Figure~\ref{fig:NEP_Footprint} are pointings that have already been acquired during 2018-2020 while the blue pointings are data that will be acquired by $\sim$ 2025. The H20 imaging footprint can be seen as the white circles in Figure~\ref{fig:NEP_Footprint}. Due to ongoing data acquisition from both surveys, this study is focused on a 194 arcmin$^2$ area where both surveys have achieved fully reduced photometric and spectroscopic data which is just a small fraction of the entire 10 $deg^2$ NEP field.  

\subsection{Spectroscopic Data}

The VIRUS instrument on the HET gathers spectra using a fiber integral field unit (IFU) system, which has a spectroscopic resolution of 4.7 to 5.6 \AA\ (resolving power of 750-950) \citep{Hill_2021}. VIRUS has a wavelength coverage of 3500-5500 \AA\, allowing for optimal \lya\ detection between redshifts 1.9 -- 3.5. 
Each pointing in Figure~\ref{fig:NEP_Footprint} consists of a maximum set of 78 VIRUS IFUs with each IFU covering a 51$^{\prime\prime}$ x 51$^{\prime\prime}$ area of the sky.

VIRUS did not start out with 78 IFUs and IFUs were continuously being added to the VIRUS instrument since 2017. During the time of our observations, there were 67 active IFUs. Each IFU has 448 1.5$^{\prime\prime}$-diameter fibers, which have a 1/3 filling factor. The standard observational strategy employs three small dithers to fill the fiber gaps, yielding fully-filled IFUs.  However, the VIRUS focal plane has an IFU filling factor of 1/4.5.  To fill in the inter-IFU gaps, the TESLA survey performs four additional pointing maneuvers so each shot consists of four 3-dither pointings which result in a maximal set of $\sim$ 300 IFUs for each pointing.  The total area covered for each shot is roughly 18' x 18' (covering the entire focal plane except the central hole, where the remaining HET instrument feeds are located). 

As TESLA uses an observing strategy that is very similar to the Hobby-Eberly Telescope Dark Energy Experiment (HETDEX) survey, a wider area survey with VIRUS, we are able to leverage the HETDEX data reduction pipeline to reduce the TESLA data \citep{HETDEX_Reduction, HETDEX}. Some notable features of the HETDEX reduction pipeline are that it: 
1) reduces and calibrates the data, 2) finds emission line(s) in the spectrum, and 3) attempts to classify any line it detects using the approach from \citet{Bayesian_line_detection} in an automated emission line classification scheme called ELiXer \citep[][Davis et al. submitted]{Elixer_Paper, Farrow_2021, HETDEX_Reduction}. The end result of the data reduction is a fully flux-calibrated, PSF-weighted spectrum.

The full methodology for data reduction, line extraction, and detection implemented by the HETDEX reduction pipeline can be found in \citet{HETDEX_Reduction} but we give a brief summary of the main components here. The HETDEX reduction pipeline finds possible emission line(s) by performing a grid search in the spectral and spatial resolution elements of each IFU. They use all the spectral and spatial resolution elements associated with each IFU. When searching for emission lines, they extract a spectrum from every fiber in the IFU, and attempt to fit an emission line weighted by the corresponding PSF at every spatial and spectral location in the data cube. The reduction pipeline uses a step size of 0.5$^{\prime\prime}$ in the spatial dimension and 8 \AA\ in the spectral dimension \citep{HETDEX_Reduction}. The pipeline extracts spectra at each location and attempts to fit a Gaussian profile on any emission lines found, keeping sources that have a S/N $>$ 4.0 and a \sqrchi\ less than 3.0 (from the Gaussian fit). The pipeline then performs a secondary grid search around each source that passed the first round of cuts, using a 5$^{\prime\prime}$ x 5$^{\prime\prime}$ box with a spacing of 0.15$^{\prime\prime}$. The second search has a more stringent requirement for line detection and requires that the emission line have a S/N $>$ 4.8 and \sqrchi\ $<$ 1.2. If an emission line satisfies these criteria they find the coordinate and extract a PSF-weighted spectrum at that position. The extraction performed here is done using the fibers closest to the coordinate of interest. The algorithm uses a weighted sum across each wavelength bin with the fiber trace as the centroid and the fiber profiles as the weights. All of the line information, spectra, and observing information are then stored in a master source catalog as described in \citet[][submitted]{Cooper_2022}. %

\begin{figure*}
    \centering
    \includegraphics[width = .95 \textwidth]{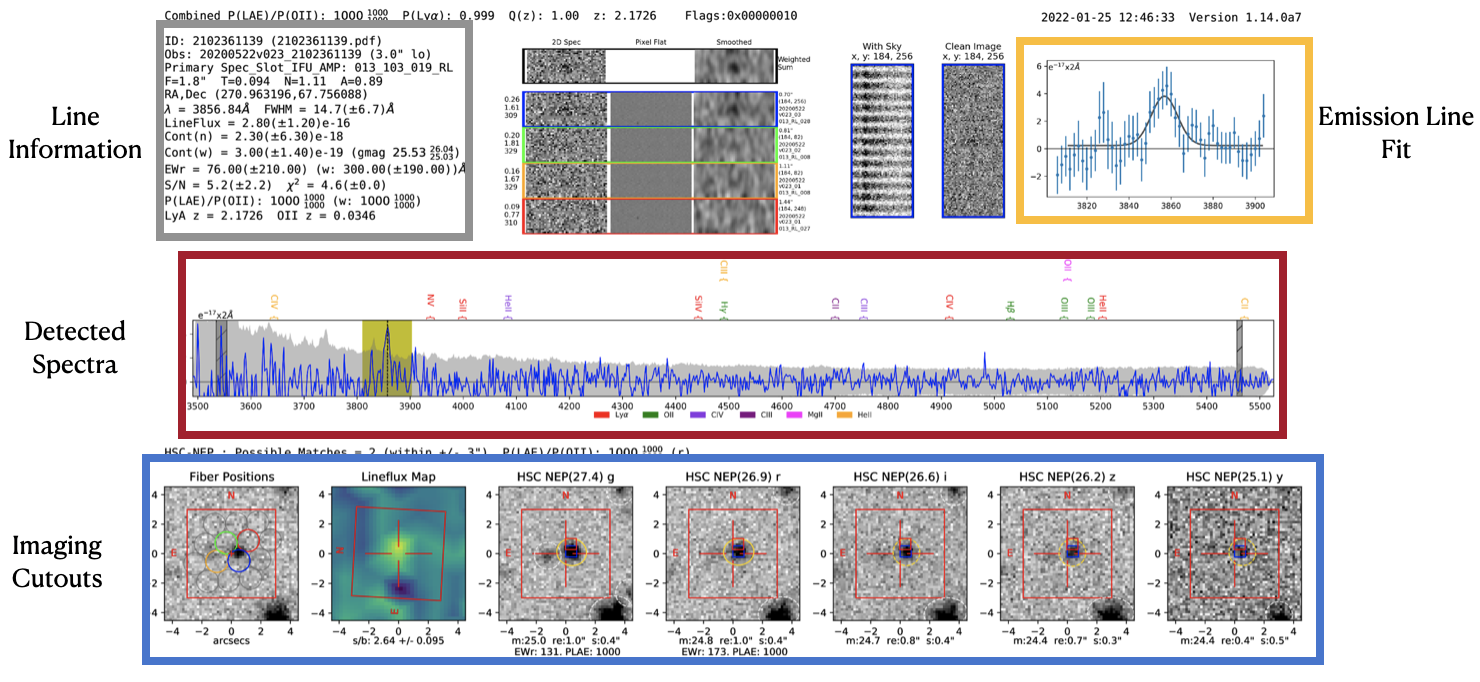}
    \caption{An ELiXer report of an LAEs in our final sample.
    \textbf{Line Information:} A summary of the detected emission line. It contains the coordinate of the extracted spectrum, the observed wavelength of the emission line, line flux measurements, the equivalent width of the line, and the redshifts of the source assuming the emission line is \lya\ or [\ion{O}{2}]. 
    \textbf{Detected Spectra:} The one dimensional PSF weighted spectrum of the source. The highlighted region shows the emission line found by the line detection algorithm. One thing to note is that the VIRUS instrument cannot reliably detect a continuum for the typical brightness of LAEs between $z =$ 2 -- 3.5.
    \textbf{Imaging Cutouts:} A series of imaging cutouts centered on the detection coordinate. The LineFlux map shows the spatial extent of the emission line near the emission line center. The rest are different imaging cutouts of different bands. 
    \textbf{Emission Line Fit:} A zoom-in on the detected emission line with the best fit Gaussian shown as the solid line in the image. With the range of information at our disposal, we could say that this line is \lya\ at z = 2.17 with a high degree of confidence coming from the galaxy in the blue box in the imaging panels.}
    \label{fig:elixer}
\end{figure*}

We use the latest available data release at the time of this work, which is the internal HETDEX Data Release 2 version 3 (HDR2.1.3) \citep[][submitted]{Cooper_2022}, which includes data from the TESLA field. HDR2.1.3 has the emission line detection catalog along with Emission Line eXplorer (ELiXer) reports and derived quantities used in the ELiXer reports \citep[][]{Elixer_Paper}. The ELiXer reports include important emission line quantities and quality control of the emission line and VIRUS during the time of observation for each detection. We show an ELiXer report of one of the LAEs in our pilot study to show the wealth of information available to us in Figure~\ref{fig:elixer}. It includes useful diagnostics that one can use to visually inspect the source if any doubts arise, such as the raw two-dimensional charged-coupling devices (CCDs), sky-subtraction cutouts, quality of the fit, imaging cutouts, and fiber locations. The main piece of information to highlight for our work is the \lya\ emission strength, which can be seen as one of the values in the gray box in the upper left, as well as the redshift for this source, also shown in the gray box next to the \lya\ redshift. 

To find LAEs we took the entire catalog of emission line detections and filtered it by right ascension (RA) and declination (DEC) to find detections that overlapped with our available photometric footprint (the zoom-in in Figure 1). The final number of emission line detections in the region of our pilot study resulted in 2726 detections. While this is a big number, the true number of Ly$\alpha$ lines is much smaller.  First, this emission line catalog goes down to a S/N$=$4.8, and the spurious fraction at this S/N can be high \citep[][submitted]{HETDEX_Reduction, Cooper_2022}. Second, the wavelength range covered by VIRUS probes low redshift galaxies, most prominently [\ion{O}{2}] (3727 \AA) at $z <$ 0.5. Thus, a way to determine whether an emission line detection is real and whether it is \lya\ or [\ion{O}{2}]3727 was needed. This is something that deep imaging and photometry can help us uncover.

\subsection{Photometric and Imaging Data}

To accurately determine emission line classification and perform SED fitting we use imaging data
from the Hawaii 20-deg$^2$ survey (\hto; \citet[][in prep]{Mcpartland_2022}). The H20 survey is actively acquiring and reducing deep optical imaging data from the Subaru Hyper SuprimeCam of the 10 $deg^2$ NEP field (as part of the Cosmic Dawn Survey \citet[][in prep]{Mcpartland_2022}, their imaging footprint can be seen as the big white circles in Figure~\ref{fig:NEP_Footprint}. \hto\ reaches $5\sigma$ depths for point sources in the following bands: $g$ = 27.5, $r$ = 27.5, $i$ = 27, $z$ = 26.5, $y$ = 25.0 AB magnitudes and they have access to \textit{Spitzer}/IRAC photometry from the SPLASH Survey which reaches depths in bands 3.6$\mu m$ = 24.8 and 4.5$\mu m$ = 24.9. The y-band data is archival data and was not taken by the \hto\ team, the y-band data used in the catalog was reduced by \citet[][]{Oi_2020}. The \textit{Spitzer}/IRAC photometry is further discussed in \citet[][]{Moneti_2022}. Since this study is done at the edge of the H20 survey the magnitudes differ slightly from the average survey region and are actually $g$ = 27.1, $r$ = 26.7, $i$ = 26.4, $z$ = 26, $y$ = 24.9 and \textit{Spitzer}/IRAC 3.6$\mu m$ and 4.5$\mu m$ = 25.

The H20 survey implements a new method to extract photometry for sources using a forced flux modeling method called \textsc{The Farmer} \citep[][Weaver et al. in prep]{Farmer_Method}, which uses the profile fitting tools provided by \textsc{The Tractor} \citep{Tractor_a, Tractor_b, Farmer_Method}. The methodology of how \textsc{The Farmer} operates is as follows: 

\begin{enumerate}
    \item Sources are detected using the pythonic source extractor (SEP, \citealt{SEP}) on a $r$, $i$, and $z$ \texttt{CHI\_MEAN} combined image built with SWARP \citep{swarp}.
    \item Detected sources are modeled simultaneously using input from $r$, $i$, and $z$ bands adopting either a point source or extended source determined based on goodness-of-fit.
    \item Photometry is measured by optimizing the PSF-convolved model on each $g$, $r$, $i$, $z$, $y$ band independently, leaving flux as a free parameter.
\end{enumerate}

\noindent A more detailed explanation of how \textsc{The Farmer} finds sources, and models the fluxes of sources can be found in \citet{Farmer_Method}. The H20 photometric measurements are further described in \citet[]{Zalesky_2023}.

The biggest strength that this photometric catalog has is that the forced modeling approach is able to acquire accurate fluxes from faint and/or blended sources, which would be difficult to get using standard aperture photometry. Another advantage of using the H20 imaging and photometry is that it helps us to disentangle the ambiguity between an emission line being \lya\ or [\ion{O}{2}]3727. The deeper imaging allows for more accurate continuum estimates than we can get from the VIRUS spectra, as no continuum is detected in the spectra of most LAEs (see spectra in Figure~\ref{fig:elixer}), making equivalent width measurements more robust. We use the equivalent width of the line as a criterion to separate a \lya\ emitter from an [\ion{O}{2}] emitter as we expect the equivalent width of [\ion{O}{2}] emitters to be low when compared with \lya\ emitters \citep[][]{Gawiser_2006}.

\begin{figure*}
    \centering
    \includegraphics[width = .95\textwidth]{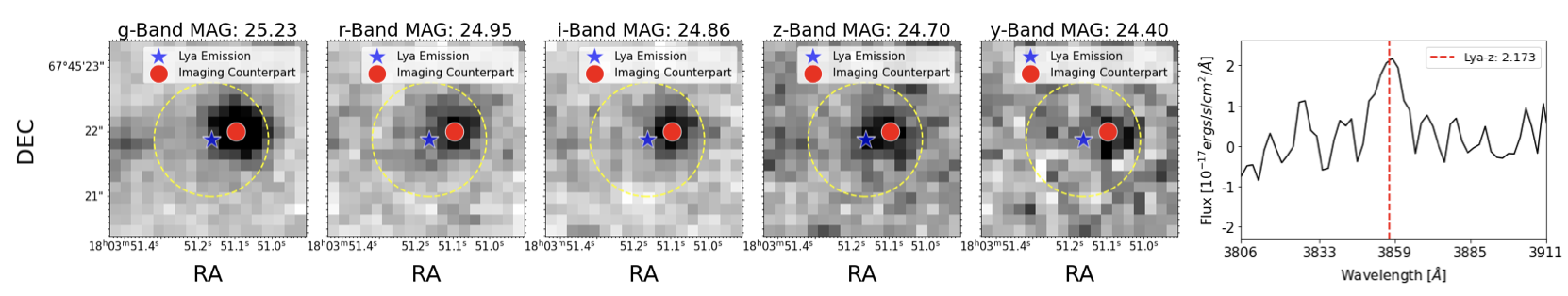}
    \caption{HSC H20 multi-band cutouts of an LAE in our sample which was found using a search radius of 1$^{\prime\prime}$ and a radially dependent S/N threshold. The HETDEX detection position is shown by the blue star, the imaging counterpart position in red, and the yellow circle outlines the HETDEX seeing full-width half maximum (FWHM) which was for this source 1.76$^{\prime\prime}$. This example highlights our robust methodology to find continuum sources for emission line detections.}
    \label{fig:hsc_cutout}
\end{figure*}

\section{Methodology} \label{sec:methods}

Having uncovered 2726 emission line detections in the 194 square arcminute region we needed a method to match imaging counterparts to the emission lines and to accurately classify the emission lines. In this section, we outline our approach to generate our sample of LAEs. 

\subsection{Counterpart Identification}

Since the HETDEX pointings are unbiased pointing, having no preselection of sources, identifying imaging counterparts to the emission line detections is not straightforward. Some factors that lead to this are that the emission line finding code, in the reduction pipeline, can find an emission line that is spatially offset from the barycenter of the galaxy. As the emission line extraction uses a Point Spread Function (PSF) weighted extraction which can impart a spatial shift to the detected line away from the barycenter of the imaging counterpart. There is also the issue with astrometric uncertainties trying to cross-match between the emission line detection catalog to the \hto\ photometric catalog. As a result of these minor astrometric uncertainties, we implement a search radius of 1$^{\prime\prime}$ centered on the emission line's spatial location to find continuum counterparts. The barycenter of the galaxy is found using The Farmer, which uses SEP \citep[][]{SEP} and is left as a free parameter in The Farmer's modeling of the source \citep[][]{Farmer_Method}. This radius was chosen following earlier HETDEX work with deep {\it Hubble Space Telescope} imaging, finding that when imaging counterparts were detected, they were almost always within 1$^{\prime\prime}$ of the detected position \citep{mccarron2022}.

We use the imaging position in the $r$-band as a prior for the emission line position by implementing a radially-dependent signal-to-noise (S/N) threshold of the emission line. Emission lines that were closer to the imaging counterpart are more likely to be real and were allowed to be at, slightly, lower S/N. We consider imaging sources that fell in these regimes a match to the emission line detection: 

\begin{itemize}
    \item Separation between 0$^{\prime\prime}$ -- 0.5$^{\prime\prime}$ and an emission line S/N $>$ 5 
    
    \item Separation between 0.5$^{\prime\prime}$ -- 0.75$^{\prime\prime}$ and an emission line S/N $>$ 5.5 
    
    \item Separation between 0.75$^{\prime\prime}$ -- 1$^{\prime\prime}$ and an emission line S/N $>$ 6 
\end{itemize}

Figure~\ref{fig:hsc_cutout} shows a galaxy from our 194 square arcminute region that was found using this selection method. The blue star shows the emission line spatial location found by the PSF-weighted algorithm which is based on the line flux contribution from the nearby fibers. The red circle shows the center of the continuum counterpart and the yellow circle outlines the full-width at half-maximum (FWHM) of the seeing during the VIRUS observation. With this method, we are able to find roughly 500 continuum counterparts to the 2726 emission lines. The rest of the detections are a combination of spurious detections and real detections that had no significant imaging counterpart detected. This is seen in \citet[][submitted]{Cooper_2022} as roughly half of the LAEs in the publicly released HETDEX Source Catalog do not have an imaging counterpart to an $r$-band sensitivity limit of 26.2 mag. To reduce contamination due to stars and bright low-redshift galaxies, we apply an $i$-band magnitude cut of $i >$ 20 AB Mag, reducing our galaxy counterpart sample to a total of 242 galaxies. 

\subsection{Emission Line Classification}

When the HETDEX reduction pipeline finds only a single emission line it is unclear whether the line is \lya\ or an [\ion{O}{2}]3727 emission line (most other potential emission lines would result in two detections; e.g., H$\beta$ and [\ion{O}{3}]5007). The deeper H20 imaging helps us break the degeneracy because a galaxy emitting [\ion{O}{2}]3727 will have a smaller equivalent width than an LAE which is measurable via deep imaging (as the spectroscopy is often too shallow to detect the continuum, see Figure~\ref{fig:elixer}, \citet[][]{Gawiser_2006}).

To identify the emission line as \lya\, [\ion{O}{2}]3727, or some other emission line we designed a decision tree algorithm that combines spectroscopic and photometric information by fitting a source with the SED-fitting code EAZY \citep{EAZY}. This code uses all available photometry and compares it to a series of galaxy templates, allowing nonlinear combinations of any number of provided templates. We use the included EAZY template set, “tweak\_fsps\_QSF\_v12\_v3,” which is based on the Flexible Stellar Population Synthesis code \citep{conroy10}. This template set has been corrected (or “tweaked”) for systematic offsets observed between data and the models. To this template set, we add six additional templates spanning bluer colors than the FSPS models, as described in \citet{Larson_2022} which found that these improve the accuracy of photometric redshift fits for bluer galaxies. 

The new templates were created using stellar population models created with BPASS \citep{eldridge09}. To generate bluer rest-UV colors than the initial set of FSPS templates, they selected BPASS templates with low metalicities (5\% solar), young stellar populations (log stellar ages of 6, 6.5, and 7 Myr), and were inclusive of binary stars. They added an additional set of these models which also include nebular emission lines derived with CLOUDY \citep{ferland17} using a high ionization parameter ($\log$ $U = -$2), low metallicity (0.05 Z$_{\odot}$), and with nebular continuum emission. \citet[][]{Larson_2022} provides sets of these templates without \lya\ (for high-redshift galaxies), with reduced \lya\ emission (either 1/3 or 1/10 of that produced by CLOUDY), and with full \lya\ strength. For this paper, we use the 12 tweak FSPS template and also include the 3 BPASS only templates plus the 3 BPASS + CLOUDY templates with \lya\ reduced to 1/10 as this is akin to a 10\% escape fraction which previous studies have found to be the average \lya\ escape fraction of LAEs (see \citet{Pucha_2022, Weiss_paper, Sobral_2017}).

We then run three different EAZY fits for each source with different parameters as described below: 

\begin{enumerate}
    \item Fixed-redshift model assuming the emission line is Ly$\alpha$ (\lya\ Fit)
    \item Fixed-redshift model assuming the emission line is [\ion{O}{2}]3727 ([\ion{O}{2}] Fit)
    \item A free redshift parameter model (Free-z Fit)
\end{enumerate}

\begin{figure*}[t]
    \centering
    \includegraphics[width = .9 \textwidth]{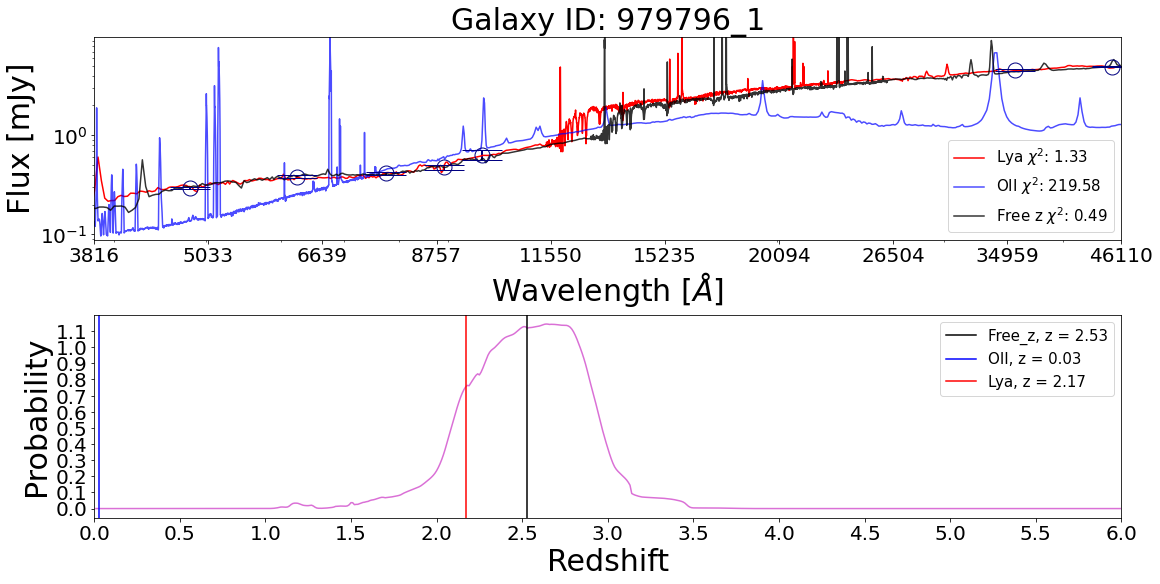}
    \caption{\textbf{Top}: This plot shows the best fit EAZY models and their corresponding \sqrchi\ for the three models implemented in the emission-line classification decision tree: 1) fixed-z assuming the line is \lya, 2) fixed-z assuming the line is [\ion{O}{2}], and 3) a free-z parameter model. \textbf{Bottom}: The free-z model's redshift posterior distribution shows that the SED fit for this source heavily favors a high-z solution, making it more likely that this line is \lya. Both plots work together in the decision tree to come up with a classification of the source, and both pieces of information shown here point to this source being an LAE. A more detailed look into the decision tree algorithm can be seen in Figure~\ref{fig:flow_chart}.}
    \label{fig:Decision_SED}
\end{figure*}

To quantify how well an EAZY model was able to fit the photometric data we investigate the \sqrchi\ values reported by EAZY. The \sqrchi\ tells us how well the EAZY-generated galaxy SED was able to fit the data while accounting for errors in the photometric measurements. A lower \sqrchi\ indicates that the model was able to fit the photometric data well. We compare the best-fit model's $\chi^2$ between the \lya\ and [\ion{O}{2}] fits against each other and check if one is significantly better than the other. The first criterion we used to determine a good fit was a \sqrchi\ $< 10$, this criterion was determined after visually inspecting a sub-sample of sources to provide a good EAZY fit while allowing for some uncertainty since the {\it Spitzer}/IRAC photometry could have been biased by nearby sources, which leads to a higher \sqrchi\ value. The second criteria we used was a \sqrchi\ difference between the \lya\ model and the low-redshift model, ie $\Delta\chi^2  = |\chi^2_{Ly\alpha} - \chi^2_{low\_z}| > 10$. In conjunction with criteria one, we are able to classify the emission line as either \lya\ or [\ion{O}{2}]. A source selected using this criterion can be seen in Figure \ref{fig:Decision_SED}, we can see that the \lya\ fit is much better than the [OII]-fit and additional support can be seen in the redshift distribution preferring a high-redshift solution.

When the $\chi^2$ values between the [\ion{O}{2}] and \lya\ fits are low and comparable (ie: $\chi^2 < 10$ for either the \lya\ or [\ion{O}{2}]-fits \textbf{and} $\Delta \chi^2 < 10$) we resort to using the free-z fit's redshift posterior distribution, P(z), the equivalent width of the line assuming \lya\ (\EW), calculated using Equation \ref{eq:EW} with $f_{line}$ being the integrated flux of the best fit Gaussian using a window of $\pm$ 2$\sigma$ and $f_{cont}$ being the flux of the source in the $r$-band from the photometric catalog, and the $r$-band magnitude. 

\begin{equation}
    W_{Ly\alpha, rest} =\frac{f_{line}}{f_{cont}(1+z_{Ly\alpha})}
    \label{eq:EW} 
\end{equation}

First, we integrate the P(z) in three regimes:

\begin{enumerate}
    \item \textbf{Low-z}:   0 $< z <$ 0.6, checks for [\ion{O}{2}] and [\ion{O}{3}] Emission
    \item \textbf{Middle-z}: 0.6 $< z <$ 1.9, checks for other emission lines (such as [\ion{Mg}{2}])
    \item  \textbf{High-z}: z $>$ 1.9, checks for \lya\ Emission
\end{enumerate}

Once we find the integrated P(z) regime that has the highest probability density, we examine the \EW\ and magnitude to corroborate the classification. For the low-z regime, we check that the \EW\ $< 20$ \AA\ and $r$-band magnitude $< 23$, for the high-z regime we check that the \EW\ $> 20$ \AA\ and $r$-band magnitude $ > 23$. We use the 20 \AA\ threshold for LAEs as this value has been found by \citet[][]{Hu_1998, Bayesian_line_detection, Shapley_2003} to be a good diagnostic of selecting high-redshift \lya\ emitting galaxies. 

To classify a source as neither \lya\ or [\ion{O}{2}], we require that the free-z parameter model's $\chi^2$ be substantially better than both the [\ion{O}{2}]-fit and \lya -fit. We use $\Delta \chi^2 > 30$ as our threshold for being a better fit, and the integrated P(z) between redshifts .6--1.9 be higher than 0.65, making most of the probability density be in this redshift range. We employ this strict $\chi^2$ difference due to the free-z model having an extra free parameter which makes it easier to fit the photometric data. If both of these conditions are met we classify the emission line as not belonging to \lya\ or [\ion{O}{2}] and it does not get included in our final sample of LAEs. 

For sources where the EAZY fits were poor, defined as \sqrchi\ $>$ 40 for the \lya\ \textbf{and} [\ion{O}{2}] fit, we ran the exact same models but excluding the IRAC bands. We did this because we uncovered that in many of these cases the IRAC photometry heavily biased the fits and increased the \sqrchi\ of each model fit. This is likely due to poorly deblended IRAC photometry, such that the IRAC fluxes make it incompatible with the optical SED. Due to removing the IRAC photometric data points we cannot reliably constrain the redshift posterior using the optical $g$, $r$, $i$, $z$, $y$ bands. We decided it was best to \textbf{not} use the information from the redshift posterior distribution in our final line classification for sources that went this route in the decision tree algorithm as the P(z) is not very well constrained.  

We note that there may be times when we may have some contradictory information when it comes to various aspects of the decision tree. These can be contradictions between the \sqrchi\ classification with an expected \EW\ and/or magnitude value, or when the magnitude and \EW\ seem to contradict one another from what we expect an [\ion{O}{2}]-emitter or LAE to have. For sources where we had conflicting information, we placed them in a category where visual inspection was needed so that we can either add them to the LAE sample or remove them altogether. A flowchart outlining the most common path taken by our sources can be seen in the Appendix in Figure \ref{fig:flow_chart}.

Applying the decision tree to the 242 counterpart matches resulted in the classifications shown in Table \ref{table:Classification}, with 185 sources classified as [OII]-emitters, 8 sources requiring further analysis, 0 classified as galaxies within redshift 0.6 - 1.9, and 49 as LAEs.  We did not add any galaxies from the visual inspection sample to the final LAE sample as they all showed signs of being low-z sources, such as the source being notably spatially resolved in the ground-based imaging, ie. being more extended than the typical 1.2'' seeing, and had equivalent widths $<$ 20 \AA. We also note that it is possible that a single line detection could also be confused as [OIII]5007 when the emission line is faint, such that the [OIII]4959 and H$\beta$ lines would not be detected. Contamination by faint [OIII] emitters could contaminate only the highest redshift \lya\ emitting galaxies ($\lambda_{obs} >$ 5007 \AA; $z_{Ly\alpha} > $ 3.1) and lowest SN sources. We investigated the five galaxies in our sample with $z_{Ly\alpha} > $ 3.1 to explore the potential for contamination by very low-redshift [OIII] emission.  We find these sources have \EW\ $> 30 $ \AA,  $r$-band magnitude $ > 23$, and are compact in the imaging (sources being smaller than the typical seeing FWHM in Figure \ref{fig:43_LAEs2}), making it very unlikely that [OIII]5007 contamination is significant.

\begin{table}
    \centering
    \begin{tabular}{|c|c|}
         \hline
         Classification & Number of Sources \\
         \hline
         Further Analysis & 8\\
         \hline
         [\ion{O}{2}]-Emitters & 185\\
         \hline
         Middle-z Sources & 0\\
         \hline
         LAE & 49\\
         \hline
    \end{tabular}
    \caption{Final emission line classification on the 242 imaging counterparts using our decision tree algorithm. We see that a substantial amount of the sources have been classified as [\ion{O}{2}]-emitters and only 49 are classified as LAEs. With visual inspection we found AGNs and duplicate emission line detection in the LAE sample which reduces our final LAE sample from 49 to 46.}
    \label{table:Classification}
\end{table}

After applying our decision tree algorithm, we did a final round of visual inspection to ensure that everything classified as an LAE was correctly classified. We were checking to see if the emission line was narrow, which would indicate a noise spike in the spectrum. We also checked for broad emission lines this feature would indicate the source is an AGN. After a quick round of visual inspection, we uncovered two AGNs based on their spectral shape, broad emission line features, and signature of broad CIV emission that got misclassified as LAEs, which we removed from the LAE sample. There were also 2 emission lines that were duplicate detections, due to multiple VIRUS observations of the same source, and we kept the observation with the highest S/N. After the visual inspection, we end up with a final sample of 46 high-confidence LAEs. 

\subsection{SED Fitting Routine}

\begin{table}
    \centering
    \begin{tabular}{ |c|c|c|c| } 
        \hline
        Parameter & Prior & Bounds & Units \\ 
        \hline\hline
        Age of Galaxy & Uniform & 0, 13 & Gyr \\ 
        \hline
        $\tau$ & Uniform & 0.3, 10. & Gyr \\ 
        \hline
        $M_{formed}$ & Uniform & $10^6$, $10^{12}$ & $M_{\odot}$ \\ 
        \hline
        $Z$ & Uniform & $10^{-5}$, 2 & $Z_{\odot}$ \\ 
        \hline
        $A_V$ & Uniform & 0, 2 & mag \\ 
        \hline
        $\log U$ & Uniform & -4, -2 & - \\
        \hline
    \end{tabular}
    \caption{Table showing the BAGPIPES fit parameters passed into our delayed-tau SFH model. We selected the bounds on physical arguments and the priors to be as un-informative as possible.}
    \label{table:bp_param}
\end{table}

We measure global galaxy properties using SED fitting techniques on our final sample of 46 spectroscopically-confirmed LAEs with imaging counterparts. We fit these 46 LAEs with the Bayesian SED fitting code \bagpipes\ to acquire posterior distributions of global galaxy properties \citep{Bagpipes}. \bagpipes\ uses the MultiNest sampling algorithm to efficiently sample the posterior distribution of a complicated multi-dimensional parameter space \citep{MultiNest}. To generate galaxy SED models, \bagpipes\ uses the 2016 spectral templates from \citet{BC_Models}, assumes a \citet{Kroupa_IMF} initial mass function, and includes nebular modeling by implementing nebular lines from the 2017 version of CLOUDY \citep{ferland17}. More information about \bagpipes\ implementation and modeling can be found in \citet{Bagpipes}.

To account for a variable star formation history (SFH), such as an increasing or declining SFH, we fit our LAEs with a delayed-tau SFH model, which allows for both possibilities. Table \ref{table:bp_param} shows the fit parameters and the corresponding priors used in our \bagpipes\ fits. We fix the redshift prior to the spectroscopic redshift found by the reduction pipeline, removing one source of uncertainty. 

\begin{figure}%
    \centering
    \includegraphics[width = \columnwidth]{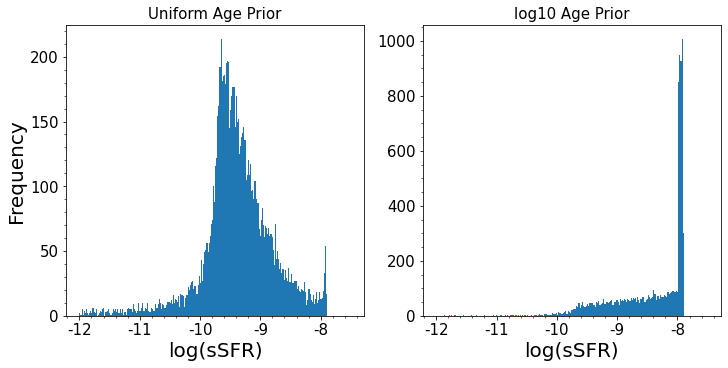}
    \caption{\textbf{Left:} This is the prior distribution of the sSFR after setting the prior on age to be uniform. We can see that this prior is not restrictive in what values it is able to explore, covering a range of log$_{10}$(sSFR) of -12 all the way to -8. This figure highlights that the choice of BAGPIPES priors drastically affects the explored quantities for the posteriors. \textbf{Right:} This is the prior distribution on the sSFR after setting the prior on age to be uniform in $\log_{10}$ (ie: uniform in $\log_{10}$(age)). Such a prior seems to prefer sources that have a large sSFR. This is due to galaxies with this prior being pushed toward smaller ages as a uniform in $\log_{10}$ prior is a $\frac{1}{x}$ distribution. To be as least constraining as possible we set all the priors in our BAGPIPES fits to be uniform in linear space.}
    \label{fig:priors}
\end{figure}

When it comes to performing Bayesian inferences, the choice of priors is key to the inference one makes about the posterior. We cannot impose a prior on quantities such as SFR or sSFR, as there is no direct SFR prior in \bagpipes\, but the choice of prior on the age of the galaxy or $\tau$ can end up affecting this derived property. To illustrate this we show in Figure~\ref{fig:priors} the impact of two different age priors on one galaxy and their effects on the sSFR. Each plot in Figure~\ref{fig:priors} has 1000 random draws of the sSFR from the same galaxy, but we change the prior on the age. The right plot in Figure~\ref{fig:priors} has a prior on the age of the galaxy that is uniform in $\log_{10}$ space (ie: $\log_{10}(Age(t_1, t_2)$), where the left plot is a prior on the age of the galaxy that is uniform in linear space (ie: U($t_1, t_2$), where $t_1$ and $t_2$ correspond to the bounds in the age on Table \ref{table:bp_param}. The age prior that is uniform in $log_{10}$ space reduces the likelihood of low sSFR solutions and preferentiates sSFR values near log$_{10}$(sSFR) of $-$8. While the prior is uniform in linear space is more spread out, encompassing both high and low sSFR solutions. We find this happens because \bagpipes\ computes the SFR over the last 100 Myr and using SFR = $M_{*}/100 Myr$ we get that sSFR = $1/100 Myr$ which results in the sSFR pile up near $-$8. After performing this diagnostic on our priors, the priors listed in Table \ref{table:bp_param} were chosen in such a way that the data would be driving the results and to minimize our effect on biasing the SED fitting in any way.

Another precaution that we took to ensure accurate SED fits is for sources at $z \geq$ 2.2 we removed the $g$-band flux from the photometric fitting (by setting the flux to zero and the uncertainty to $10^{99}$) as the $g$-band flux will contain the \lya\ emission at these redshifts, and we want to avoid strong \lya\ emission biasing the fits in our SED fitting routine. After using \bagpipes\ to fit all 46 galaxies we looked at their \sqrchi\ values to assess the quality of the fits. We noticed that three sources had high \sqrchi\ values, with \sqrchi\ $> 100$, indicating a bad fit, and as a result, we removed these sources from our final sample bringing the final number of LAEs to 43. We show 3'' x 3'' cutouts and zoom-ins on the \lya\ emission of these 43 LAEs in the Appendix in Figure \ref{fig:43_LAEs2} as well as general source information of these 43 LAEs in Table \ref{table:Line_Info}.

\subsection{Equivalent Width Estimates}

The main focus of our work, beyond developing our sample selection methodology, is to study how SED-derived galaxy properties relate to the strength of \lya. To this end, we use the equivalent width of \lya\ as our proxy for the \lya\ strength to compare galaxy properties against. Using the equivalent width allows one to study the strength of \lya\ in a way that is normalized to the galaxy's continuum. To quantify the \EW\ on the 43 LAEs we used the \bagpipes\ generated spectrum to determine the continuum level near the \lya\ emission line. We utilized the \bagpipes -explored parameter values to generate model spectra for every realization and compute the continuum over the range of 1225-1250 \AA\ in the rest frame.  For every sample generated by \bagpipes\ we compute the average flux in this window using Equation \ref{eq:cont_est}:

\begin{equation}
    \label{eq:cont_est}
    f_{avg} = \frac{\int_{1225}^{1255} \lambda T_{\lambda} f d\lambda}{\int_{1225}^{1255} \lambda T_{\lambda} d\lambda}
\end{equation}

\noindent where we assumed $T_{\lambda} = 1$ in this range, f is the \bagpipes\ modeled flux, and  $f_{avg}$ is our continuum estimate for a given realization.

\begin{figure}
    \centering
    \includegraphics[width = .43\textwidth]{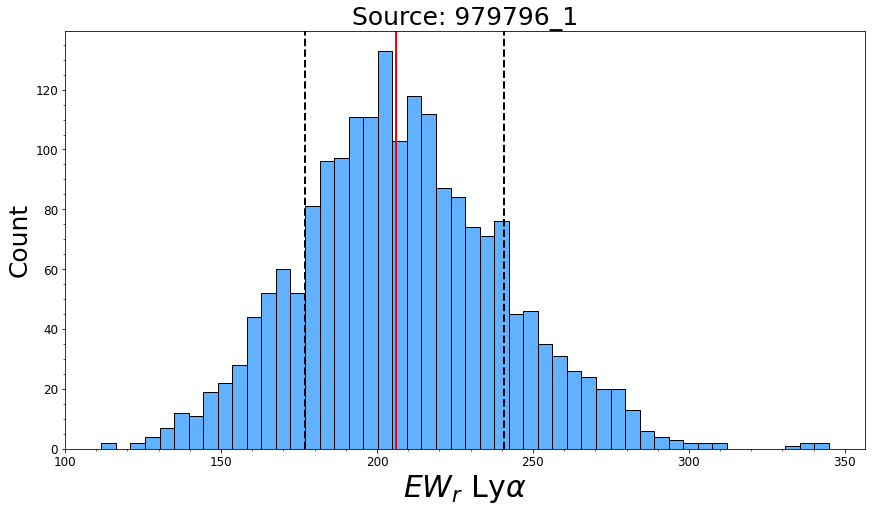}
    \caption{The equivalent width posterior distribution for one LAE in our sample from our Monte Carlo method, where we use $\sim$1000-2000 \bagpipes\ realizations to compute the continuum near the line, and perturb the HETDEX emission-line flux values within their uncertainties. The solid line shows the median of the distribution with the dashed lines showing the 16th and 84th percentile. In this particular source this distribution is centered at a large \EW\ value with most of the distribution being above 100 \AA.}
    \label{fig:EW_MC}
\end{figure}

We computed $f_{avg}$ for every realization that \bagpipes\ explored to get a distribution of continuum estimates for every LAE. To get a corresponding distribution of flux values we drew fluxes from a Gaussian distribution centered at the line flux value found by the reduction pipeline and used a standard deviation that was equal to the flux error found by the  data reduction pipeline. We drew as many realizations of the line flux as was needed to match the \bagpipes\ realizations ($\approx$ 1000-2000) and divided the distribution of fluxes by the distribution of \bagpipes\ continuum estimate to get a final distribution of the \EW\ for each source. Lastly, we converted the observed \EW\ to rest frame by dividing by 1+$z_{Ly\alpha}$. We report back the median value from this distribution as the \EW\ values presented in the paper and we use the absolute difference between the 16th and 84th percentile to the median as the lower and upper uncertainty on our \EW\ estimate. A plot of the \EW\ distribution from one LAE using this method can be seen in Figure~\ref{fig:EW_MC}. We note that due to the large uncertainty in our flux measurement, attributed to its low S/N, this naturally gets propagated to the final \EW\ we compute.

\subsection{Computing Correlations}

To explore correlations between a galaxy property and the \EW\ strength we need a way to quantify correlations robustly. We use the Spearman-r correlation coefficient, as this uses a rank-based approach and is invariant to any transformation of the \EW\ (ie: galaxy property vs \EW\ and galaxy property vs $\log_{10} (W_{Ly \alpha})$ measures the same correlation). To quantify the uncertainty in our measured correlation we used a Monte Carlo (MC) approach. For every galaxy we have a ``chain" of 1000 values that \bagpipes\ has explored, we use all 43 galaxies at the same step in the chain to compute the Spearman-r correlation coefficient between a galaxy property and \EW. This involves making a 43 x 1000 matrix, where each row is a different galaxy, and the columns are the posterior values of a given galaxy property at each time step in the chain. We took each column and implemented a bootstrap method to compute correlations. We randomly drew a random set of 43 galaxy property values, with replacement, in each column and acquired the corresponding 43 \EW\ values to compute the Spearman-r correlation coefficient. The bootstrap method was used so that our correlation was not biased by any outliers in the data due to our small sample size. This method gave us a distribution of Spearman correlation coefficients, we report back the median as our fiducial correlation and quantify the upper and lower uncertainty as the absolute difference between the 16th and 84th percentile to the median. To visually plot the uncertainties of our correlations as we were computing the Spearman correlation coefficient, we also fit a line to the data for every realization in the bootstrap procedure. The spread of the linear fits in Figure~\ref{fig:EW_Corr} highlight the 16th and 84th percentile of these line fits which capture the spread of our correlation coefficients.

\section{Results} \label{sec:results}

The methodology outlined in \S \ref{sec:methods} resulted in a final sample of 43 LAEs, which we fit with \bagpipes\ to derive their global galaxy properties. Here we outline some key results as well as initial correlations with galaxy properties and \lya\ emission strengths. We note that because of the small sample size, we do not expect to make any robust correlations between a galaxy property and the \lya\ line strength with this sample, but rather these results are a proof of concept for a future study with a larger TESLA LAE sample.

\begin{figure*}[t]
    \centering
    \includegraphics[width = .9 \textwidth]{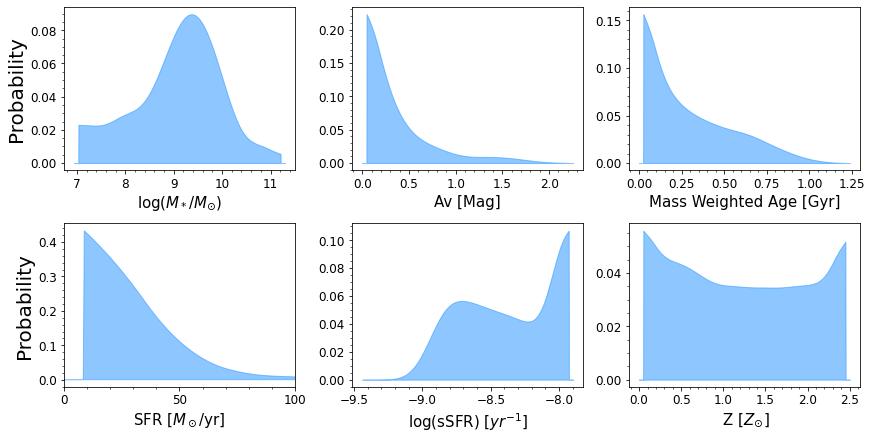}
    \caption{The stacked posterior distribution for all 43 spectroscopically-confirmed LAEs in our sample. This plot shows that the most probable stellar mass for our sample of galaxies is centralized around $ 10^9 - 10^{10}$ $M_{\odot}$, and that most of our LAEs likely have low dust attenuation and modest SFR (SFR $ < 50 M_{\odot}/yr$). We do notice what seems to be a bi-modality in the $\log_{10}(sSFR)$ distribution with some LAEs preferring really high $\log_{10}(sSFR)$ of $-$8, indicating really young ages, and another peak at around $-$8.75, indicating slightly older stellar populations. The metallicity is not well-constrained, which is expected given the lack of constraining power of photometric data on the metallicity. Overall the trends seen in our stacked posterior show that our LAEs seem to be in line with what has been shown in the literature.}
    \label{fig:All_Gal_Dist}
\end{figure*}

First, we explored the average properties of our sample. We took the posterior distribution for every galaxy property in our final sample, namely stellar mass, dust attenuation, mass-weighted age, SFR, log(sSFR), and metallicity, and made a normalized histogram of their distributions. The distributions were smoothed out using a Gaussian filter using a value of 1.5, in units of the data passed in, for the standard deviation of the Gaussian kernel to transform the discrete bins into a more continuous distribution. Figure~\ref{fig:All_Gal_Dist} shows the stacked posterior distributions of global galaxy properties of all 43 LAEs.

The stellar mass posterior distribution of our 43 LAEs has a peak at around $10^9-10^{10} M_{\odot}$ (upper left on Figure~\ref{fig:All_Gal_Dist}). Most of the galaxies also show signs of having low dust attenuation as shown in the upper middle panel of Figure~\ref{fig:All_Gal_Dist} with most of the probability distribution preferring dust attenuation below 0.5 magnitudes. The sample has a wide range in SFR with most of the sample being within 8 -- 75 $M_{\odot}$/yr, but with a strong preference towards lower SFR. The specific star formation rate (sSFR) distribution of the 43 LAEs shows signs of a bimodal distribution with some LAEs preferring high log$_{10}$(sSFR) near $-$8 $yr^{-1}$, indicative of a younger galaxy population, and another high probability peak near more moderate sSFR at log$_{10}$(sSFR) = $-$8.7 $yr^{-1}$. The metallicity in Figure~\ref{fig:All_Gal_Dist} shows that our photometric data are (as expected) unable to constrain this parameter, showing an almost uniform distribution for this parameter. Despite our inability to constrain metallicity, we leave the metallicity as a free parameter in the SED fits so that the uncertainty in this parameter is propagated to the other derived properties.

We next explore potential correlations between \EW\ and galaxy properties. In Figure~\ref{fig:EW_Corr} we uncover that the strongest correlations are between \EW\ and stellar mass, and \EW\ and star formation rate, each at the $\sim 1\sigma$ level. These have modest anti-correlations with a Spearman correlation coefficient of -0.$34_{-.14}^{+.17}$ for stellar mass and -0.$37_{-.14}^{+.16}$ for SFR. Figure~\ref{fig:EW_Corr} shows the regression plots where the uncertainty in our correlation coefficients is represented by the shaded region in the plots. These figures highlight the power TESLA has in studying \lya\ emission from galaxies, but with our small sample, we are limited in our interpretation of the results. An expanded data set is needed to explore these correlations at a higher significance.

\section{Discussion} \label{sec:discussion}

There have been many papers and studies outlining the properties of LAEs at various redshifts. Here we attempt to put our sample into context with previous studies in the literature to see where our sample stands. 

\subsection{Galaxy Properties}

We can see in the upper left panel of Figure~\ref{fig:All_Gal_Dist} that the 43 TESLA LAEs cover a wide range of stellar masses ranging from $10^7 - 10^{11} M_{\odot}$, with the peak of this distribution near $\log_{10}(M_*/M_{\odot}) \approx$ 9.5. This is a wide dynamic range in the stellar masses of LAEs, implying that TESLA has the capability of finding LAEs with a wide range in stellar masses. There has been other work in the literature that has found stellar masses of LAEs at these redshifts. For example, \citet{Guaita_2011} has compared other redshift 2-3 LAEs in the literature and found that LAEs stellar mass range from log$_{10}$(M/$M_{\odot}$) = 8.12 -- 9.83, which is contained within the peak of our stellar mass distribution.   \citet{Gawiser_2007} studying 162 LAEs at redshift 3.1 measured a stellar mass of stacked LAEs to be  1 x 10$^9 M_{\odot}$, which lines up with the peak of our stellar mass distribution in Figure~\ref{fig:All_Gal_Dist}. \citet{Hagen_2016} carried out a similar analysis of LAEs using the HETDEX survey and found that their unbiased search of LAEs also spans a wide range of stellar masses spanning $10^{7.5} - 10^{10.5}$ $M_{\odot}$. This points out that an unbiased search for LAEs can generate a sample that spans a large dynamic range in stellar mass, with \citet{Hagen_2016} stellar mass spanning 3 orders of magnitudes and TESLA LAEs spanning a bit over 3 orders of magnitudes, ranging from log$_{10}$($M_*/M_{\odot}$) = 7.09 -- 10.81 using the 43 LAEs in this sample. The stellar mass has been shown to be an important parameter in understanding the escape of \lya\ in galaxies. To see if this is the case for TESLA LAEs we need to expand the sample and see what the large sample statistics point to.

The combined SFR posterior distribution covers a wide range of SFR values spanning 8 -- 100 $M_{\odot}/yr$ with most of the distribution favoring low SFR, as seen by the pile-up near SFR of 8 $M_{\odot}/yr$. Despite the preference for low SFR, the TESLA survey has the capabilities of probing even higher SFR sources as indicated by the tail of the distribution going all the way to 100 $M_{\odot}$/yr. Looking into the literature we see that LAEs are a mix of both low and high SFR as indicated by different studies. On the low SFR regime (SFR $<$ 10 $M_{\odot}$/yr) we have \citet{Gawiser_2007} measuring an SFR of 2 $M_{\odot}$/yr on a stack of LAEs at redshift 3.1, \citet{LAEs_21} measures SFRs between 0.01 -- 80 $M_{\odot}$/yr with most of the LAEs contained within SFR of 0.8 -- 10 $M_{\odot}$/yr at redshift 2.1, and \citet{LAEs_2_3} measures SFRs between 0.1-- 10 $M_{\odot}$/yr for LAEs between redshifts 2.1 -- 3.1. On the high SFR regime (SFR $>$ 10 $M_{\odot}$/yr) we have the \citet{Vandels_2019} finding LAEs between 10-100 $M_{\odot}$/yr. In \citet{Hagen_2016} the peak of their SFR is near 10 $M_{\odot}$/yr with some sources near 100 $M_{\odot}$/yr, \citet{Shapley_2003} measures an SFR of 52, and 38 $M_{\odot}$/yr for systems with \lya\ absorption, an SFR of 29 $M_{\odot}$/yr for weak \lya\ emitters, and an SFR of 25 $M_{\odot}$/yr for LAEs, following our definition of \EW\ $>$ 20 \AA. \citet{Pentericci_2007} compute an SFR value of 76 $M_{\odot}$/yr for LBGs with line emission and 74 $M_{\odot}$/yr LBGs with no line emission. With all these different studies finding a wide range in SFR for LAEs it is clear that having a survey that can probe both low and high SFR is crucial. That is something that the TESLA survey has the capability of exploring once a large sample of LAEs is assembled.

The top middle plot in Figure~\ref{fig:All_Gal_Dist} is the dust attenuation posterior distribution of the 43 TESLA LAEs and it demonstrates that low dust attenuation is preferred, with most of the distribution having the highest probability near 0.1 magnitudes. Other studies such as \citet[][]{LAEs_21, Vandels_2019, Trainor_2016, Trainor_2019}, find evidence that LAEs tend to have on average low dust attenuation. \citet[][]{Vandels_2019}, for example, found that their LAE sample has a dust attenuation E(B-V) values between $10^{-3} - 10^{-1}$, which lines up with the peak of the TESLA dust distribution near low values of $A_V$. The fact that many studies are finding LAEs exhibiting low dust is an indication that low dust is an important feature that is vital for the escape of \lya\ emission, as less dust would allow more \lya\ flux to escape the galaxy. It is worth noting that there is a non-zero probability of sources in our sample having really high dust attenuation with the probability distribution going all the way to 2 magnitudes in Av. \citet[][]{Fink_2009} explains that a dusty galaxy can have strong \lya\ emission if the specific ISM geometries allow for more open channels for \lya\ to resonantly scatter and escape the galaxy while minimizing the interaction with dust. Having more high-dust LAEs would be interesting sources to look into and further our understanding of the connection between \lya\ emission and dust. For that, we will need an expanded data set with a wide range of dust properties to see how dust affects \lya\ emission strength, which TESLA can achieve once complete.

The middle panel in the bottom row of Figure~\ref{fig:All_Gal_Dist} shows that there is a preference for high values of sSFR, with a high probability density near log$_{10}$(sSFR) of $-$8. The distribution also shows a secondary set of high probability near log$_{10}$(sSFR) = $-$8.7. \citet[][]{Hagen_2016} studied LAEs at $z =$ 2 in the HETDEX Pilot Survey, and found their sample of LAEs preferentially have a high sSFR around log$_{10}$(sSFR) $\leq$ $-$8, consistent with our probability distribution. Other studies show this trend for LAEs as well, such as \citet[][]{Nakajima_2012} who found that the LAEs they studied, when stacked and averaged, had an average sSFR near log$_{10}$(sSFR) $\approx$ -7.5. \citet[][]{Gawiser_2007} who studied LAEs at redshift 3.1 also uncovered that their sample of LAEs had a rather high sSFR of log$_{10}$(sSFR) = -8. To interpret this result we can turn to the interpretation offered by \citet[][]{Castro_2006} where they outline that high values of sSFR indicate the tracing of the youngest star-bursting galaxies. The timescale is also important here as a high value of sSFR is indicative of a younger stellar population which we see reflected in the mass-weighted age plot in the top right of Figure~\ref{fig:All_Gal_Dist}. Our sample of 43 TESLA LAEs seems to show that high sSFR values are preferred for LAEs. We would need an expanded data set to see if the trend holds true for more LAEs with a wide range of galaxy properties.

The top right plot in Figure~\ref{fig:All_Gal_Dist} demonstrates that the highest probability density of the 43 LAEs is located at very low mass-weighted ages. Other studies such as \citet[][]{Gawiser_2007, Pentericci_2007, Kornei_2010, Fink_2007, Fink_2009} have found evidence that LAEs show signs of having a young stellar population. These studies seem to indicate that observable \lya\ emission from galaxies require them to have ongoing or have undergone recent star formation making it so that these young stellar populations dominate the light from a galaxy. A recent study by \citet[][]{Pucha_2022} found corroborating results where the LAEs in their sample had ages less than 1 Gyrs with a median value at $10^7$ yrs. When analyzing our 43 TESLA LAEs we find that LAEs seem to have young ages, indicative of a young stellar population that has enough high mass stars around to produce a vast amount of \lya\ emission. More study is needed to see if this holds true for an expanded data set with a wide range of ages and galaxy properties.
\linenumbers\relax

\begin{figure*}
\centering 
  \includegraphics[width=0.95\textwidth]{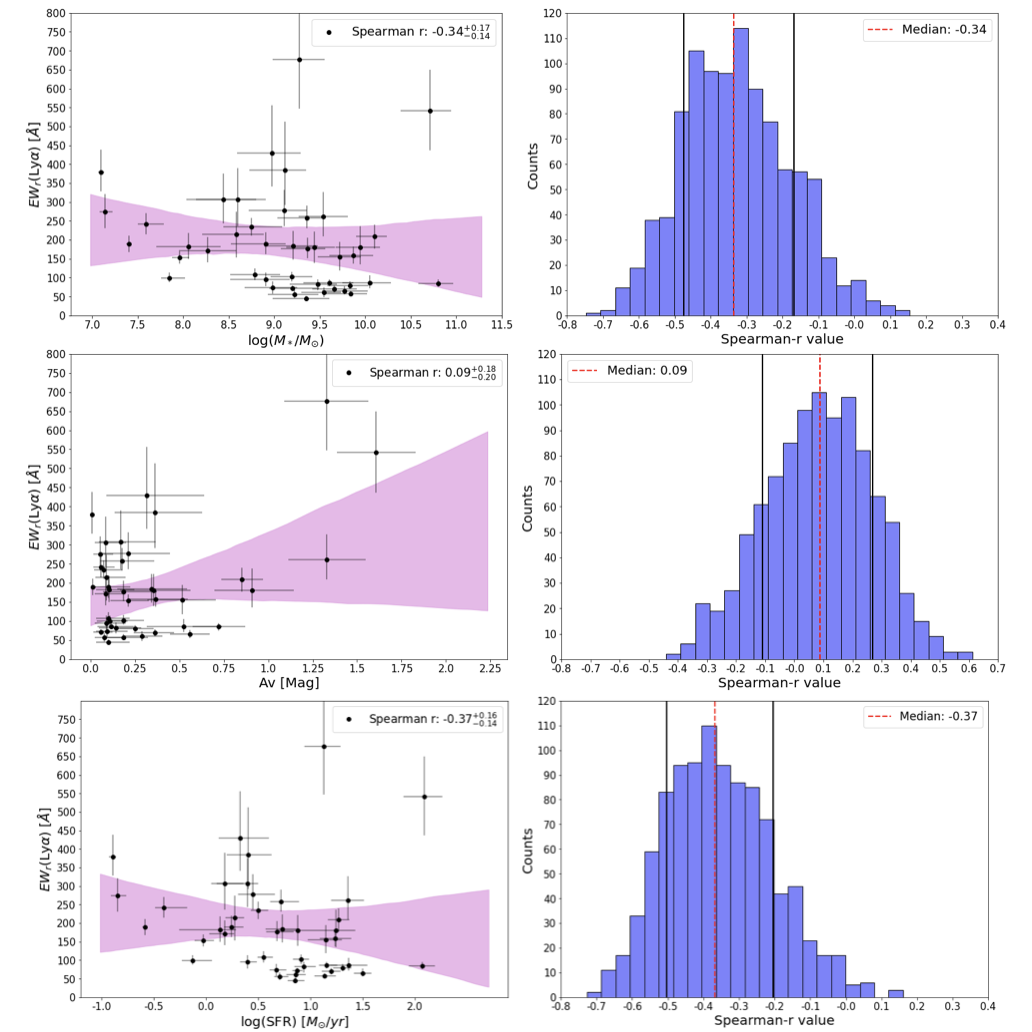}

\caption{\textbf{Top: \EW\ vs Stellar Mass:} We show a scatter plot between \EW\ and the stellar mass of the 43 LAEs in our sample. We see that there is a slight anti-correlation with a value of $-$0.34. The uncertainty on our correlation shows that the slight anti-correlation is significant at only the $1\sigma$ level. \textbf{Middle: \EW\ vs Av:} We see here that there seems to be no correlation between \EW\ and Av with most of the correlation distribution being near zero. \textbf{Bottom: \EW\ vs SFR:} The strongest anti-correlation in our sample is between \EW\ and SFR as seen in the leftmost plot but we do note that this anti-correlation is still only significant at the  $1\sigma$ level. These plots show that the galaxies with the strongest \lya\ emission come from low-mass, low SFR, and dust-poor galaxies. The blue bands on the left plot were computed by performing linear fits on a random set of 43 data values, with replacement, and repeating this 1000 times and getting the 16th and 84th percentile from the linear fits. The rightmost plot shows a histogram of the correlation values with the 16th, and 84th percentiles shown as the solid black vertical lines and the median as the dashed red vertical lines.} 
\label{fig:EW_Corr}
\end{figure*}

\subsection{Correlations}

\textbf{\EW\ vs Mass:} We find that one of the most correlated galaxy properties with \EW\ is stellar mass with a Spearman correlation coefficient of $-0.34^{+0.17}_{-0.14}$. This can be seen in Figure~\ref{fig:EW_Corr} as we can see a downward trend shown by the data. Our bootstrap method explained in section \ref{sec:methods}, uncovered that this correlation is only significant at the $1\sigma$ level, which is seen as the shaded area in Figure~\ref{fig:EW_Corr}. The shaded area shows that, within the constraints placed by our data, we can have a clear anti-correlation, no-correlation or correlation. We find that this anti-correlation has been seen in studies such as \citet{Trainor_2019} where they find a slight anti-correlation between stellar mass measuring a Spearman correlation of $-$0.15 between \EW\ and stellar mass. Other works \citep[see.][]{Vandels_2019, Jones_2012} also find an anti-correlation measuring a Spearman correlation of -0.34 in the \citet{Vandels_2019} study, and \citet{Jones_2012} stating that lower mass galaxies have stronger \EW\ implying that a negative correlation exists between \EW\ and stellar mass. Another study that sees this effect is the study of \citet{Oyarzun_2016, Oyarzun_2017} where they find clear evidence of an anti-correlation existing between stellar mass and \EW. The anti-correlation found in all these studies and more can be explained from a dust perspective, since low-mass galaxies will typically have less dust \citep[eg.][]{Brinchmann_2004, Zahid_2013}. As a result, low-mass galaxies have less \lya\ being absorbed by dust within the galaxy making it likely that more \lya\ is able to escape. Contrast this to if you have a more massive galaxy, these galaxies will have more dust and the chance of \lya\ being absorbed by dust within the galaxy is higher. The galaxy will have less \lya\ emission escaping relative to the continuum near the \lya\ line and thus we will measure a lower \EW. However, some studies have seen the opposite conclusion or no correlation present such as \citet{Kornei_2010} who report a Spearman correlation coefficient of .011 on their entire sample and .068 on their sample omitting $K_s$ non-detections, indicating a positive correlation between \EW\ and stellar mass. Their entire sample allowed them to explore low \EW\ sources (\EW\ $< 20$) as well as \lya\ absorption, whereas our work focused on high \EW\ (\EW\ $> 20$) sources which can explain the discrepancies. \citet{Hathi_2016} measured in their sample of LAEs a Spearman correlation coefficient between \EW\ and stellar mass of $-$0.08, indicating little to no correlation. In summary, we find a tentative correlation at the $1\sigma$ level between \EW\ and stellar mass. This implies that stellar mass could be used to predict the strength of \lya\ emerging from galaxies. 

\textbf{\EW\ vs Dust:} The middle plot of Figure~\ref{fig:EW_Corr} shows the regression plot for \EW\ vs $A_V$ and we do not find a significant correlation, computing a Spearman correlation coefficient of 0.09$^{+0.18}_{-0.20}$. We can say with the 43 TESLA LAEs that we do not find a definitive correlation between dust and \EW. Studies such as \citet{Oyarzun_2017} studying galaxies between redshifts 3.5 -- 4.5 also find that there seems to be no correlation between \EW\ and dust (see their figure 11). Contrasting this to other studies such as \citet{Vandels_2019} measuring a Spearman correlation coefficient of $-$0.43, \citet{Shapley_2003} seeing an anti-correlation in their Figure 14, \citet{Kornei_2010} seeing an anti-correlation in their Figure 8. All these studies seem to find evidence of an anti-correlation existing between \lya\ emission and dust. This is found in other studies such as \citet[][]{Trainor_2016} who measured a slight anti-correlation in their sample of LAEs with a Spearman correlation coefficient of -0.15. \citet{Trainor_2019} also measures in their sample of 637 star-forming galaxies an anti-correlation between dust and \EW\ computing a Spearman correlation coefficient of -0.23. One explanation as to why we do not see a clear anti-correlation could be that we selected the TESLA LAEs to have an \EW\ $>$ 20 \AA\ so we are missing the low \EW\ sources that could be responsible for driving the \EW-dust correlations. Since all these studies that have seen an anti-correlation have all been seen with low-dust systems and with low \EW, our sample selection potentially excluded these low-\EW\ systems and can explain why we do not see an anti-correlation. Another explanation may be attributed to our dynamic range in $A_V$ where we probe both high dust and low dust systems. The previous studies that have seen an anti-correlation have looked at very dust-poor systems. For example, \citet[][]{Vandels_2019} found E(B-V) of their sources to be between $10^{-3}- 10^{-1}$, \citet[][]{Shapley_2003} found E(B-V) between -.2 - .5, and \citet[][]{Kornei_2010} sources spans 0 - .6, where they find a median E(B-V) of 0.17. Due to our huge dynamic range in dust we are not as sensitive to correlations that can be driven by the low dust systems due to the impact of the high dust sources influencing our derived correlations. The multivariate analysis of \citet{Runnholm_2020} also shows that dust and \EW\ are anti-correlated and can be used as a way to predict the escape of \lya. Since many studies have found a trend to exist between \EW\ and dust, it is clear that we need to have an expanded data set with a wide range of dust and \EW\ to say definitively if there is a trend within the TESLA LAEs.

\textbf{\EW\ vs SFR:} In Figure~\ref{fig:EW_Corr}, we show that the \EW\ is anti-correlated with SFR. We compute a Spearman correlation coefficient of $-.37_{-.14}^{+.16}$, a slight anti-correlation at the 1$\sigma$ level. This is in line with others who have seen this trend (e.g., \citet{Kornei_2010} see their Figure 8; \citet{Hathi_2016} measuring a Spearman correlation coefficient of $-$0.23, \citet{Trainor_2019} measuring a Spearman correlation coefficient of $-$0.17 from their SED SFR and \citet{Oyarzun_2017} see their figure 9) but is a feature that is not seen in \citet{Vandels_2019} where they state they did not find any correlation in their study, citing a Spearman correlation coefficient of 0.14. One explanation for this discrepancy is that the \citet{Vandels_2019} focused on emission lines only where \citet{Kornei_2010}, \citet{Trainor_2019} and \citet{Hathi_2016}, looked at both emission and absorption of \lya. It is worth noting that our method of selecting sources was also using \lya\ in emission just like the \citet[]{Vandels_2019} study but we compute a negative correlation between \EW\ and SFR. One explanation for these discrepancies could be the \EW\ values explored by the two studies. Our study has on average higher \EW\ with a majority of our sources having \EW\ $>$ 100 \AA. whereas the \citet[][]{Vandels_2019} compute most of their sources having \EW\ less than 100 \AA. To explain the anti-correlation we can follow the same arguments as the stellar mass and \EW\ correlation. SFR and stellar mass are closely related by the star-forming main sequence relation and, as a result, we would expect to uncover a correlation in SFR if there exists one for stellar mass. The latest study by \citet[][]{Pucha_2022} seem to show that there is no correlation between their LAEs and SFR computing a Spearman correlation coefficient of $-$0.02. To really hone in on whether a correlation exists between \EW\ and SFR a larger sample is definitely needed. Once complete TESLA will expand the number of LAEs and provide us with a wider range of galaxy properties and allow us to hone in on correlations between \EW\ and a given galaxy property. 

\subsection{Equivalent Width Distribution}

Various studies \citep[eg.][]{Shapley_2003, LAE_low_study1, LAEs_21, Blanc_2011, Santos_2020, Hashimoto_2017} have found that the number of LAEs as a function of \EW\ is well represented by an exponential distribution of the form:

\begin{equation}
    N(W_{Ly\alpha}) = Ae^{W_{Ly\alpha}/W_0}
\end{equation}

Where $W_0$ is the e-folding parameter ranging from study to study but hovers from 50-100 \AA\ at $z =$ 2--3. 

Due to our small LAE sample, the choice of bin width can drastically alter the resulting shape of the \EW\ distribution. To ensure that our \EW\ distribution was not biased by bin size we implement the technique known as "pseudo"-binning as outlined in \citet{Psudoebin}. The methodology for our "pseudo"-binning is as follows:

\begin{enumerate}
    \item Make a list of bin-centers, we used an array starting from 5 - 686 \AA\, in increments of 5 \AA.
    \item At each bin center make 1000 random bin widths, drawn by sampling a uniform distribution. (We drew from a uniform distribution between 1-10 \AA).
    \item Make the "pseudo"-bins using the bin-center $\pm$ bin width value drawn from the uniform distribution
    \item Perturb the \EW\ along the errors
    \item Count the number of \EW\ values that landed in the bins
    \item Repeat for all bin centers
\end{enumerate}

\begin{figure}
\centering 
\includegraphics[width = .49\textwidth]{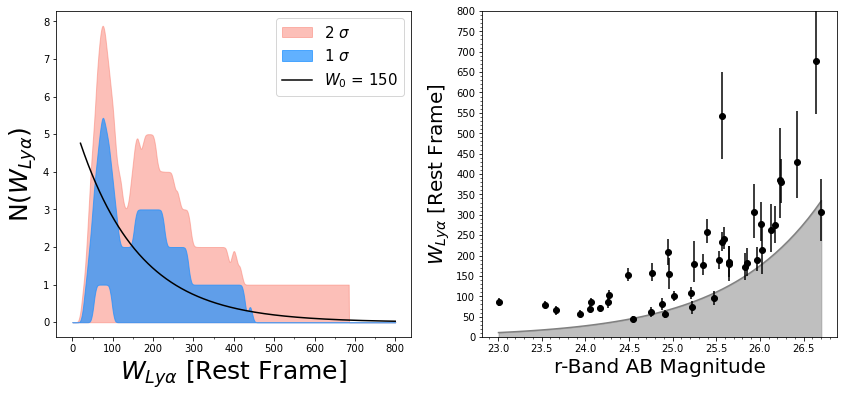}
\caption{\textbf{Left:} Plot of the \EW\ distribution of the 43 LAE sample in the pilot study. This plot shows the 1$\sigma$ and 2$\sigma$ uncertainty using our "pseudo"-binning to avoid biasing our distribution with a choice of bin width. We show for comparison an \EW\ distribution with a $W_0$ of 150, which is consistent with our distribution, going right down the middle of our 1 and 2 $\sigma$ contours. We do see a bump in our distribution near 200 \AA, which could be a by-product of our sample selection, as it may select the highest \EW\ sources, but could also be due to lower metallicities in some of our LAEs. \textbf{Right:} Our \EW as a function of magnitude. The gray shaded area shows our 5$\sigma$ \EW\ threshold. The lack of low-EW faint sources is consistent with this sensitivity curve, while the lack of high-EW bright sources appears to be real (though we are limited by the small number of bright sources)}
\label{fig:EW_Hist_EW_Mag}
\end{figure}

We use the outcome from the "pseudo"-binning to derive our \EW\ distribution and see how our \EW\ distribution compares to the \EW\ distribution of LAEs from the literature. In Figure~\ref{fig:EW_Hist_EW_Mag} we show the 1 and 2 $\sigma$ contours for our \EW\ distribution after "pseudo"-binning. We show for comparison an \EW\ distribution model with $W_0 =$ 150 \AA. We show that our LAE \EW\ distribution is consistent with an  $W_0 =$ 150 \AA\ exponential distribution capturing the excess at very high \EW\ values at \EW\ $>$ 150 \AA. We attribute this high \EW\ excess to our selection criteria in the decision tree which picks out the highest equivalent width sources. The discovery of high \lya\ equivalent width has been seen in prior studies namely \citet{Cantalupo_2012, kashikawa_2012, Santos_2020, Hashimoto_2017, Malhotra_2002}, and even seen in prior HETDEX pilot studies \citep[see][]{Blanc_2011}. \citet{Malhotra_2002} who studied LAEs at higher redshifts (z = 4.1) found a sample of very high \EW\ sources ( \EW\ $>$ 240 \AA) and they were able to explain the high equivalent width sources by invoking a top-heavy IMF and the addition of Population III stars, or by the inclusion of narrow-line AGNs. This could be one explanation for our high \EW\ sources but due to our small sample size, we cannot investigate that further. We also note that our \EW\ distribution deviates from the exponential model at \EW\ $\sim$ 40 \AA; this is understandably due to incompleteness setting in, due to our decision tree primarily only selecting sources with \EW\ $> 20$ \AA. 

In Figure~\ref{fig:EW_Hist_EW_Mag} we show the \EW\ of our sources as a function of $r$-band magnitude. We note that the highest \EW\ sources do belong to the faintest sources, with $r$-band magnitude fainter than 24.5. Figure~\ref{fig:EW_Hist_EW_Mag} also shows a gray-shaded region that outlines our equivalent width detection limit as a function of $r$-band magnitude. We computed the gray-shaded region using the median flux error of the 43 LAEs in our sample multiplying it by 5 to get the $5 \sigma$ uncertainty and used that as our flux in Equation \ref{eq:EW}. We used the $r$-band flux of each source as our continuum estimate and then used the \lya\ spectroscopic redshift to convert to rest frame \EW. The shaded region outlines sources that we will be unable to detect. Thus, with these selection effects, we are unable to acquire really faint and low \EW\ sources. We expect the lack of low \EW\ and faint sources to be consistent with the black sensitivity curve in Figure~\ref{fig:EW_Hist_EW_Mag}. We also note that we expect the lack of high \EW\ sources to be a real feature however, we are still plagued by the small number of bright sources to say conclusively \citep[][]{Ando_2006}. 
\begin{table*}
\centering
\begin{tabular}{p{.1\textwidth}p{.08\textwidth}p{.08\textwidth}p{.08\textwidth}p{.08\textwidth}p{.08\textwidth}p{.1\textwidth}p{.09\textwidth}p{.08\textwidth}p{.09\textwidth}p{.08\textwidth}}
\hline
ID &  RA & DEC &  Av  &  log$_{10}$($\frac{M_*}{M_{\odot}})$ &  SFR  &  Mass Weighted  &   $W_{Ly\alpha}$ & Redshift &  F$_{Ly\alpha}$ [$10^{-16}$]\\
\hline
Units &  Deg &  Deg & AB Mag &  & M$_{\odot}$/yr &  Age [Gyr]  &  \AA & & ergs/s/$cm^2$/ \AA \\
 \hline
979796\_1   & 270.962922  & 67.756127  & $0.86_{-0.12}^{+0.11}$ & $10.10_{-0.21}^{+0.14}$ & $18.60_{-3.78}^{+4.73}$ & $0.69_{-0.31}^{+0.25}$ & $208.24_{-31.15}^{+31.38}$ & 2.17 & 2.7\\ 
631874\_40  &  270.356110  &  67.756135 &$0.37_{-0.22}^{+0.26}$ & $9.11_{-0.34}^{+0.24}$ & $2.57_{-0.96}^{+1.68}$ & $0.48_{-0.29}^{+0.30}$ & $384.48_{-92.99}^{+128.54}$  & 2.45 & 1.97\\ 
630513\_71  &  270.368782  & 67.732457  &  $0.22_{-0.10}^{+0.11}$ & $7.96_{-0.08}^{+0.10}$ & $0.95_{-0.16}^{+0.25}$ & $0.00223_{-0.0005}^{+0.0007}$ & $152.81_{-15.26}^{-17.19}$
  & 2.15 & 4.98 \\ 
862610\_93  &  270.431767  &  67.834031 &  $0.10_{-0.07}^{+0.14}$ & $8.02_{-0.34}^{+0.36}$ & $1.26_{-0.73}^{+0.63}$ & $0.03_{-0.02}^{+0.06}$ & $182.05_{-32.83}^{+36.79}$ & 2.50 & 1.63\\ 
980044\_107 &  270.862041  & 67.758951  &  $0.57_{-0.11}^{+0.12}$ & $9.78_{-0.25}^{+0.24}$ & $32.04_{-5.56}^{+7.31}$ & $0.14_{-0.08}^{+0.19}$ & $65.39_{-10.35}^{+11.92}$  & 2.16 & 3.55\\
556041\_128 &  270.415873  & 67.671392  & $0.10_{-0.07}^{+0.11}$ & $8.97_{-0.33}^{+0.28}$ & $4.68_{-0.69}^{+1.13}$ & $0.15_{-0.09}^{+0.19}$ & $73.35_{-15.48}^{+15.69}$  & 2.89 & 1.02\\
554402\_136 &  270.399143  & 67.642269  & $0.18_{-0.11}^{+0.19}$ & $9.37_{-0.29}^{+0.19}$ & $4.67_{-0.93}^{+2.24}$ & $0.46_{-0.26}^{+0.24}$ & $176.60_{-25.45}^{+28.39}$  & 2.67 & 2.14\\
854825\_141 &  270.778197  & 67.704814  & $0.72_{-0.13}^{+0.15}$ & $10.81_{-0.21}^{+0.16}$ & $116.63_{-25.78}^{+39.05}$ & $0.55_{-0.26}^{+0.241}$ & $84.98_{-9.29}^{+9.81}$  & 2.53 & 6.93\\
854570\_142 &  270.784837  & 67.700370  &  $0.012_{-0.009}^{+0.019}$ & $7.40_{-0.02}^{+0.02}$ & $0.26_{-0.01}^{+0.02}$ & $0.0020_{-0.0003}^{+0.0003}$ & $189.47_{-22.18}^{+21.53}$  & 2.53 & 2.15\\
854780\_143 &  270.752684  & 67.704633  & $0.29_{-0.11}^{+0.13}$ & $9.55_{-0.25}^{+0.18}$ & $7.15_{-1.27}^{+1.96}$ & $0.49_{-0.28}^{+0.34}$ & $60.61_{-12.22}^{+12.26}$  & 2.19 & 1.38\\ 
726411\_157 & 270.331857  & 67.642511  &  $0.25_{-0.10}^{+0.11}$ & $9.83_{-0.25}^{+0.21}$ & $20.28_{-3.69}^{+4.50}$ & $0.30_{-0.16}^{+0.28}$ & $79.40_{-8.57}^{+10.24}$  & 2.21 & 5.58\\
727902\_162 &  270.319747  & 67.670558  & $0.18_{-0.11}^{+0.14}$ & $9.83_{-0.30}^{+0.19}$ & $13.42_{-2.74}^{+4.36}$ & $0.48_{-0.29}^{+0.33}$ & $57.40_{-6.78}^{+8.03}$  & 2.32 & 2.80\\ 
553035\_178 &  270.771389  & 67.613813  &  $1.33_{-0.24}^{+0.25}$ & $9.28_{-0.31}^{+0.27}$ & $13.33_{-4.94}^{+5.71}$ & $0.08_{-0.06}^{+0.16}$ & $676.84_{-130.12}^{+145.18}$  & 2.42 & 1.88\\
553112\_180 &  270.767107  & 67.614487  &  $0.51_{-0.19}^{+0.20}$ & $9.72_{-0.27}^{+0.22}$ & $13.76_{-4.37}^{+5.85}$ & $0.37_{-0.22}^{+0.31}$ & $155.01_{-36.42}^{+39.26}$ & 2.60 & 2.45\\
981633\_185 &  270.836433  & 67.787205  &  $0.08_{-0.06}^{+0.12}$ & $8.93_{-0.38}^{+0.24}$ & $2.47_{-0.34}^{+0.62}$ & $0.29_{-0.20}^{+0.29}$ & $95.23_{-17.28}^{+18.52}$  & 2.46 & 1.05\\
554370\_187 &  270.681597  & 67.639333  &  $0.08_{-0.05}^{+0.10}$ & $9.21_{-0.29}^{+0.27}$ & $5.08_{-0.61}^{+0.98}$ & $0.28_{-0.16}^{+0.29}$ & $56.44_{-9.70}^{+10.87}$ & 2.60 & 1.10\\
728435\_201 &  270.263506  & 67.680954  & $1.62_{-0.22}^{+0.22}$ & $10.70_{-0.32}^{+0.25}$ & $124.25_{-43.35}^{+63.10}$ & $0.42_{-0.27}^{+0.23}$ & $542.11_{-105.56}^{+107.14}$ & 2.46 & 2.97\\
727822\_202 &  270.248361  & 67.670051  &  $0.08_{-0.06}^{+0.11}$ & $8.27_{-0.32}^{+0.30}$ & $1.51_{-0.44}^{+0.32}$ & $0.07_{-0.05}^{+0.11}$ & $171.99_{-31.52}^{+35.22}$  & 2.26 & 1.67\\
552777\_210 &  270.595479  & 67.611627  &  $0.34_{-0.18}^{-0.20}$ & $9.21_{-0.31}^{0.27}$ & $5.38_{-1.64}^{+2.53}$ & $0.26_{-0.17}^{+0.30}$ & $183.22_{-35.85}^{+40.05}$  & 2.69 & 1.69\\
554682\_212 &  270.544948  & 67.645936  &  $0.11_{-0.08}^{+0.13}$ & $8.93_{-0.39}^{+0.21}$ & $1.79_{-0.29}^{+0.52}$ & $0.43_{-0.28}^{+0.29}$ & $189.92_{-28.60}^{+30.33}$ & 2.56 & 1.26\\
553018\_216 &  270.670905  & 67.614580  & $0.10_{-0.07}^{+0.13}$ & $8.79_{-0.27}^{+0.27}$ & $3.54_{-0.45}^{+0.84}$ & $0.13_{-0.07}^{+0.16}$ & $107.90_{-14.60}^{+15.58}$ & 2.43 & 1.76\\
551525\_218 &  270.774470  & 67.585961  &  $0.11_{-0.07}^{+0.13}$ & $9.60_{-0.28}^{+0.26}$ & $14.12_{-2.11}^{+3.66}$ & $0.23_{-0.13}^{+0.25}$ & $85.83_{-8.23}^{+9.13}$  & 2.82 & 3.41\\
978729\_233 &  270.964238  & 67.734907  & $0.22_{-0.15}^{+0.25}$ & $9.12_{-0.35}^{+0.26}$ & $2.82_{-0.78}^{+1.99}$ & $0.42_{-0.27}^{+0.26}$ & $277.69_{-41.61}^{+54.64}$  & 2.87 & 1.60\\
919078\_241 &  270.890146  & 67.652458  & $0.92_{-0.24}^{+0.23}$ & $9.96_{-0.27}^{+0.21}$ & $17.78_{-6.49}^{+9.38}$ & $0.52_{-0.29}^{+0.31}$ & $180.42_{-45.43}^{+56.45}$  & 2.26 & 1.78\\
982749\_253 &  270.859943  & 67.681707  &  $0.10_{-0.07}^{+0.12}$ & $7.87_{-0.11}^{+0.17}$ & $0.78_{-0.19}^{+0.43}$ & $0.003_{-0.001}^{+0.006}$ & $99.59_{-11.38}^{+14.09}$  & 2.36 & 2.22\\
859168\_255 &  270.673793  & 67.774773  &  $0.52_{-0.20}^{+0.22}$ & $10.06_{-0.32}^{+0.23}$ & $23.05_{-7.69}^{+12.18}$ & $0.50_{-0.31}^{+0.29}$ & $86.27_{-14.77}^{+19.12}$  & 2.43 & 2.60\\ 
856613\_259 &  270.807098  & 67.678839  & $0.18_{-0.12}^{+0.17}$ & $9.37_{-0.34}^{+0.20}$ & $5.28_{-1.14}^{+2.46}$ & $0.42_{-0.25}^{+0.22}$ & $257.59_{-25.95}^{+32.94}$  & 3.01 & 2.80\\ 
860820\_263 &  270.605234  & 67.801796  &$0.31_{-0.22}^{+0.33}$ & $8.99_{-0.33}^{+0.30}$ & $2.07_{-0.73}^{+1.96}$ & $0.44_{-0.28}^{+0.28}$ & $428.98_{-87.35}^{+126.67}$  & 2.69 & 1.65\\ 
855129\_280 &  270.502737  & 67.713400  &  $0.37_{-0.16}^{+0.18}$ & $9.89_{-0.29}^{+0.21}$ & $17.36_{-5.01}^{+7.40}$ & $0.43_{-0.22}^{+0.20}$ & $158.02_{-20.53}^{+23.88}$  & 3.08 & 2.97\\
727695\_283 &  270.356537  & 67.666764  &  $0.06_{-0.04}^{+0.07}$ & $7.60_{-0.09}^{+0.23}$ & $0.40_{-0.08}^{+0.34}$ & $0.002_{-0.001}^{+0.008}$ & $241.12_{-26.18}^{+29.86}$  & 2.67 & 2.63\\
981627\_306 &  270.874830  & 67.787413  & $0.10
_{-0.06}^{+0.12}$ & $9.36_{-0.38}^{+0.24}$ & $7.03_{-1.00}^{+1.69}$ & $0.27_{-0.18}^{+0.28}$ & $44.02_{-6.22}^{+7.11}$  & 2.55 & 1.22\\
859064\_310 &  270.750712  & 67.772398  & $0.07_{-0.05}^{+0.11}$ & $8.75_{-0.32}^{+0.32}$ & $3.22_{-0.55}^{+0.73}$ & $0.13_{-0.08}^{+0.19}$ & $233.68_{-23.25}^{+24.66}$  & 3.18 & 2.04\\
635175\_323 &  270.285864  & 67.692058  & $0.17_{-0.12}^{+0.22}$ & $8.58_{-0.46}^{+0.33}$ & $1.49_{-0.35}^{+0.82}$ & $0.19_{-0.14}^{+0.25}$ & $306.80_{-70.67}^{+82.15}$  & 3.12 & 0.93\\
631017\_326 &  270.358845  & 67.741362  & $0.09_{0.07}^{0.11}$ & $8.57_{0.37}^{+0.33}$ & $1.88_{-0.34}^{+0.47}$ & $0.15_{0.10}^{+0.23}$ & $213.79_{-59.98}^{+59.50}$  & 2.80 & 1.36\\
554072\_339 &  270.714379  & 67.633638  & $0.14_{-0.09}^{+0.14}$ & $9.48_{-0.30}^{+0.21}$ & $8.57_{-1.53}^{+2.70}$ & $0.30_{-0.18}^{+0.24}$ & $81.89_{-14.70}^{+13.69}$  & 3.13 & 1.42\\
858517\_363 &  270.620559  & 67.765143  &  $0.06_{-0.04}^{+0.07}$ & $9.21_{-0.26}^{+0.28}$ & $7.38_{0.68}^{1.07}$ & $0.17_{-0.09}^{+0.21}$ & $71.15_{-7.02}^{+7.44}$
  & 2.19 & 3.17\\
855444\_380 &  270.753928  & 67.716096  &$0.36_{0.12}^{+0.11}$ & $9.66_{-0.26}^{+0.23}$ & $15.79_{-3.15}^{+3.33}$ & $0.25_{-0.15}^{+0.27}$ & $69.24_{-8.42}^{+9.50}$  & 2.25 & 2.72\\
858800\_394 &  270.464903  & 67.771180  &  $0.007_{-0.005}^{+0.011}$ & $7.09_{-0.03}^{+0.04}$ & $0.13_{-0.01}^{+0.01}$ & $0.0016_{-0.0002}^{+0.0002}$ & $379.65_{-51.27}^{+57.96}$  & 2.99 & 1.48\\
858923\_396 &  270.475007  & 67.772770  &  $0.06_{-0.04}^{+0.07}$ & $7.14_{-0.06}^{+0.08}$ & $0.14_{-0.02}^{+0.03}$ & $0.0019_{-0.0006}^{+0.0009}$ & $274.99_{-43.49}^{+46.69}$  & 2.22 & 1.89\\
858780\_403 &  270.674515  & 67.682850  & $0.36_{-0.18}^{+0.21}$ & $9.44_{-0.31}^{+0.22}$ & $7.68_{-2.57}^{+4.17}$ & $0.34_{-0.20}^{+0.24}$ & $179.99_{-41.30}^{-42.52}$  & 3.19 & 1.44\\
555174\_411 &  270.683236  & 67.653930  &  $0.18_{-0.09}^{+0.12}$ & $9.18_{-0.24}^{+0.24}$ & $8.03_{-1.38}^{+1.59}$ & $0.15_{-0.08}^{+0.17}$ & ${102.12}_{-12.08}^{+13.01}$  & 2.21 & 3.49\\
637135\_432 &  270.266337  & 67.695746  &  $1.33_{-0.23}^{+0.23}$ & $9.54_{-0.28}^{+0.26}$ & $22.31_{-6.85}^{+9.76}$ & $0.09_{-0.06}^{+0.19}$ & $261.69_{-53.20}^{+64.73}$  & 2.53 & 0.99\\
554012\_461 &  270.599908  & 67.633757  &  $0.08_{-0.05}^{+0.12}$ & $8.43_{-0.40}^{+0.38}$ & $2.44_{-1.19}^{+0.66}$ & $0.06_{-0.05}^{+0.13}$ & $306.22_{-62.88}^{+68.26}$  & 3.36 & 1.89\\
\end{tabular}
\caption{Source ID, coordinates and \bagpipes\ derived properties for the 43 LAEs in our sample.}
\label{table:Line_Info}
\end{table*}

\section{Summary}\label{sec:summary}

We introduce the TESLA survey - a 10 deg$^2$ IFU spectroscopic survey over the {\it Euclid} NEP deep field which combines imaging and photometry from the \hto\ survey.  We perform a pilot study on a $\sim$ 194 $arcmin^2$ area to explore sample selection methodology and explore early correlations of LAE galaxy properties with \lya\ emission. Using a decision tree algorithm, we were able to accurately classify detected emission lines from VIRUS spectra and found a sample of 43 LAEs between z $\sim$ 1.9 -- 3.5. We use \bagpipes\ to uncover global galaxy properties of the 43 LAEs and explore correlations between galaxy properties and \lya\ equivalent width. We find the following results: 

\begin{itemize}
  \item Exploring the posterior distribution of our sample as a whole, we find that TESLA LAEs have similar physical properties as previously studied LAE samples. Our sample has the capability of probing an even wider dynamic range in stellar mass and dust. Highlighting the power of unbiased sample selections. 
  \item We find that stellar mass and SFR have the strongest correlation with \lya\ equivalent width at the $1\sigma$ level, computing Spearman correlation coefficients of -0.$34_{-0.14}^{+0.17}$ and -0.$37_{-0.14}^{+0.16}$ respectively.
  \item We show that the \EW\ distribution of our LAE sample is consistent with a typical \EW\ distribution from the literature using an equivalent width e-folding scale of $W_0 = 150$.
\end{itemize}
Due to our small sample size, we cannot make any robust claims into correlations between \lya\ emission and galaxy properties but with an expanded data set this can be mitigated. Both TESLA and the \hto\ survey are actively acquiring and reducing data which will enable the study of a larger area of the NEP field, unlocking more LAEs to study ($\sim$ 50,000 expected for the complete dataset). The larger LAE sample will enable us to study a wide dynamic range in galaxy properties, probe more of the LAE population and explore correlations more robustly than is possible with this current study. Ultimately, the findings from an expanded LAE study will inform our predictive distribution of \lya\ emission that is tied to SED-derived galaxy properties. 

When TESLA is complete and reaches the expected 50,000 LAE sample size, we will be able to robustly train our emerged \lya\ emission model at redshifts 2 -- 3.5 and calibrate it further at higher redshifts, by implementing results from higher redshift LAE studies. Thus, an increased data set will bring us one step closer to understanding the emerged \lya\ flux of galaxies based on SED-derived properties and help determine the neutral fraction during the epoch of reionization. 

\section{Software and third party data repository citations} 
Astropy \citep[][]{Astropy}, Numpy \citep[][]{Numpy}
, Pandas \citep[][]{Pandas}, Scipy \citep[][]{Scipy}, Matplotlib \citep[][]{Matplotlib}, EAZY \citep[][]{EAZY}. 

\begin{acknowledgments}

This research is based [in part] on data collected at the Subaru Telescope, which is operated by the National Astronomical Observatory of Japan. We are honored and grateful for the opportunity of observing the Universe from Maunakea, which has cultural, historical, and natural significance in Hawaii.

OCO thanks James Derry for his insight on making my big data algorithms more automated and his Python guidance as this project progressed.

OCO, SLF and GL acknowledge support from the National Science Foundation through grant AST-1908817 and NASA through grant 80NSSC22K0489.

OCO thanks the University of Texas at Austin and the Dean's Mentoring Fellowship for additional support.

The observations were obtained with the Hobby-Eberly Telescope (HET), which is a joint project 

\noindent of the University of Texas at Austin, the Pennsylvania State University, Ludwig-Maximilians-University Munchen, and Georg-August-University at Gottingen.

The HET is named in honor of its principal benefactors, William P. Hobby and Robert E. Eberly. 

VIRUS is a joint project of the University of Texas at Austin, Leibniz-Institut fur Astrophysik Potsdam (AIP), Texas A\&M University (TAMU), Max-Planck Institut fur Extraterrestrische Physik (MPE), Ludwig Maximilians-Universit at Munchen, Pennsylvania State University, Institut fur Astrophysik Gottingen, University of Oxford, and the Max-Planck-Institut fur Astrophysik (MPA).

The authors acknowledge the Texas Advanced Computing Center (TACC) at The University of Texas at Austin for providing high-performance computing, visualization, and storage resources that have contributed to the research results reported within this paper.

The Cosmic Dawn Center is funded by the Danish National Research Foundation under grant No. 140. 

We also want to acknowledge that we did this work at an institution, the University of Texas at Austin, that sits on indigenous land. The Tonkawa live in central Texas and the Comanche and Apache move through this area. We pay respects to all the American Indian and Indigenous Peoples and communities who are a part of these lands and territories in Texas.  We are grateful to be able to live, work, collaborate, and learn on this piece of Turtle Island.
\end{acknowledgments}

\newpage

\bibliography{sample631}{}
\bibliographystyle{aasjournal}

\appendix

\begin{figure*}
    \centering
    \includegraphics[width= \textwidth]{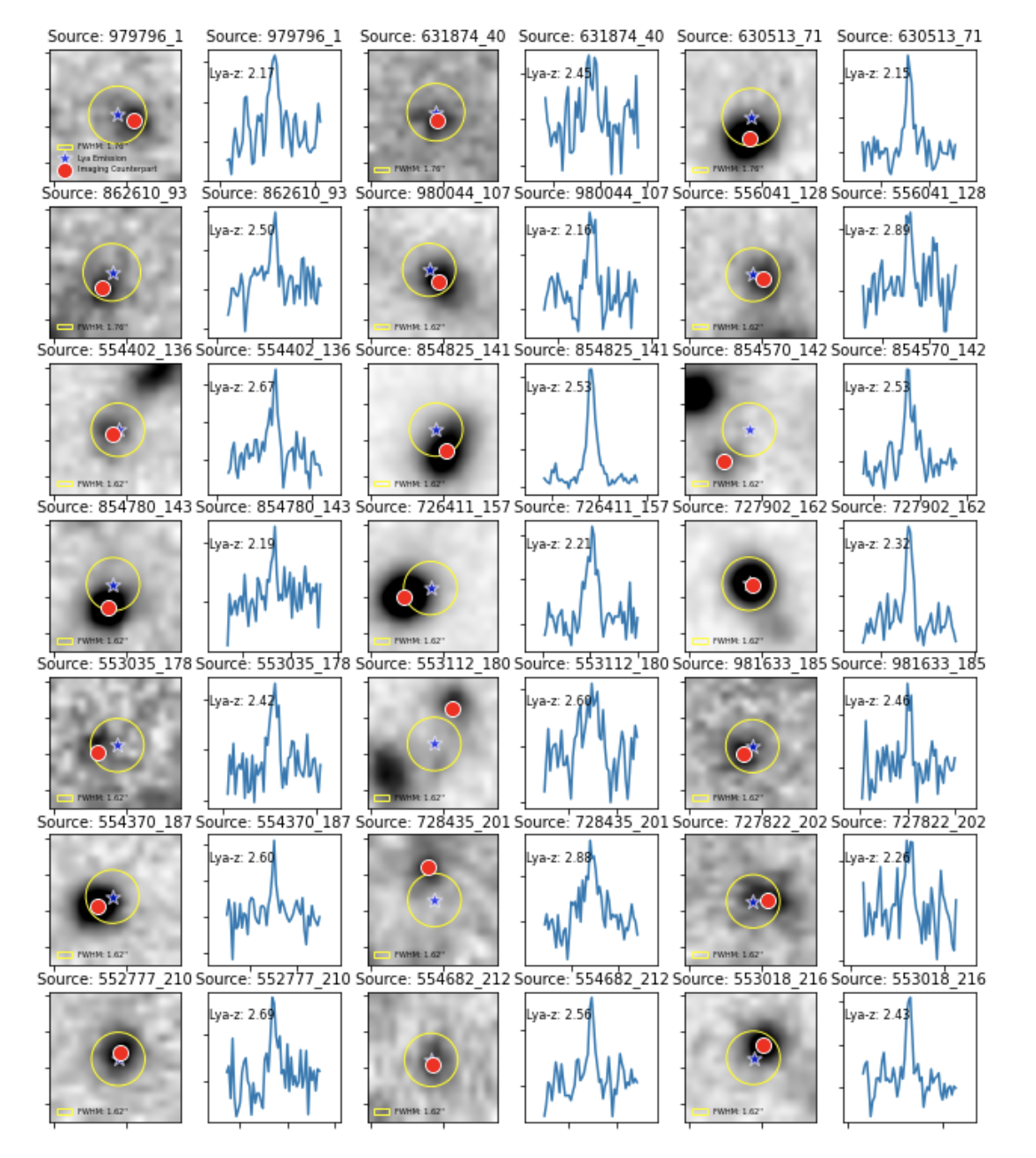}
    \label{fig:43_LAEs1}
\end{figure*}

\begin{figure*}
    \centering
    \includegraphics[width = .87\textwidth]{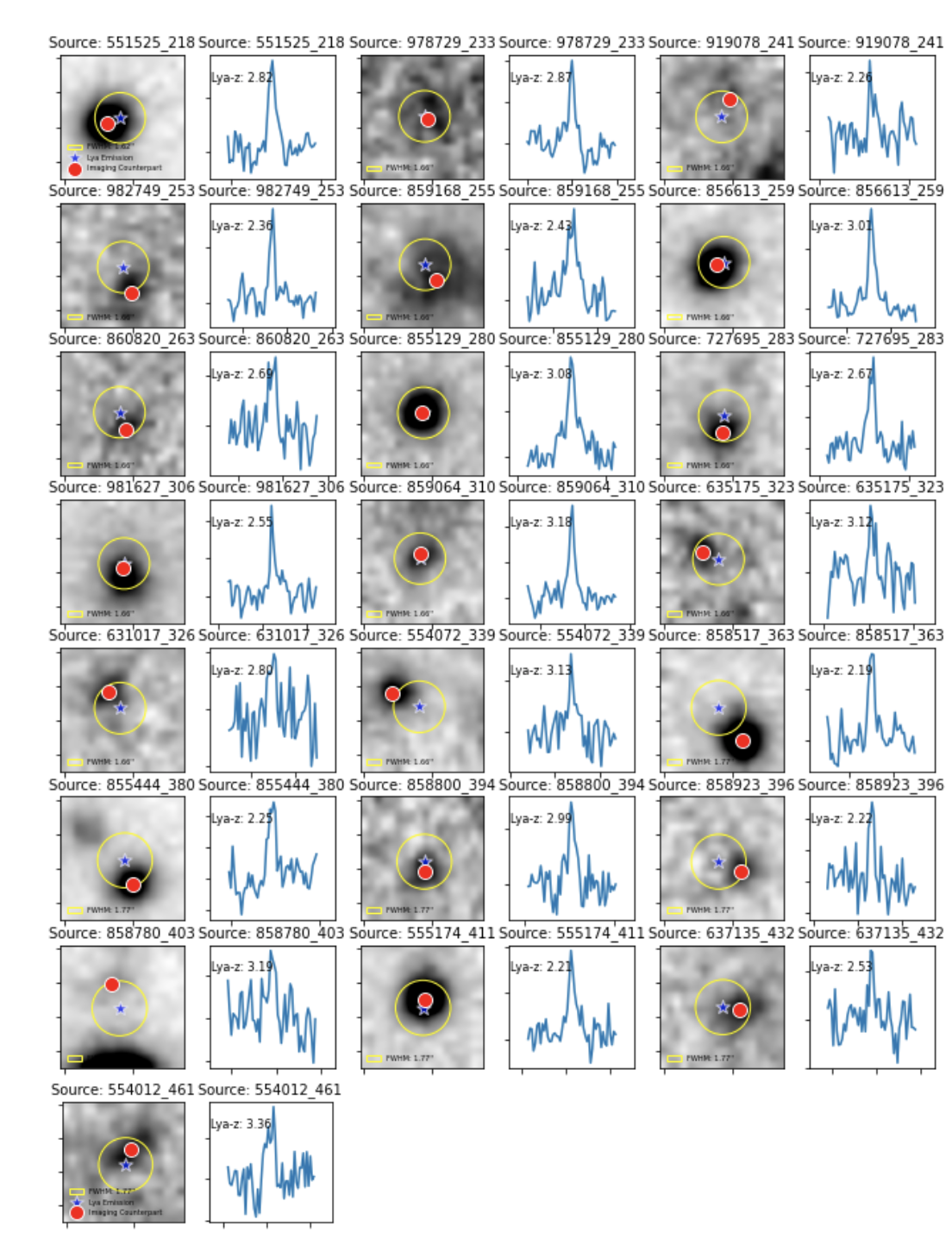}
    \caption{A 3'' by 3'' image cutout in the HSC $r$-band of the 43 LAEs in our sample. The diameter of the yellow circle in the 3'' x 3'' cutout outlines the FWHM of the seeing, in arcseconds, at the time of the observation. Alongside each cutout is the corresponding \lya\ emission line, it is centered on the observed wavelength and is zoomed in using a window of $\pm$ 50 \AA in the observed frame. The legend indicates the \lya\ spectroscopic redshift for each source. }
    \label{fig:43_LAEs2}
\end{figure*}

\begin{figure*}
    \centering
    \includegraphics[width = .87 \textwidth, height = 20cm]{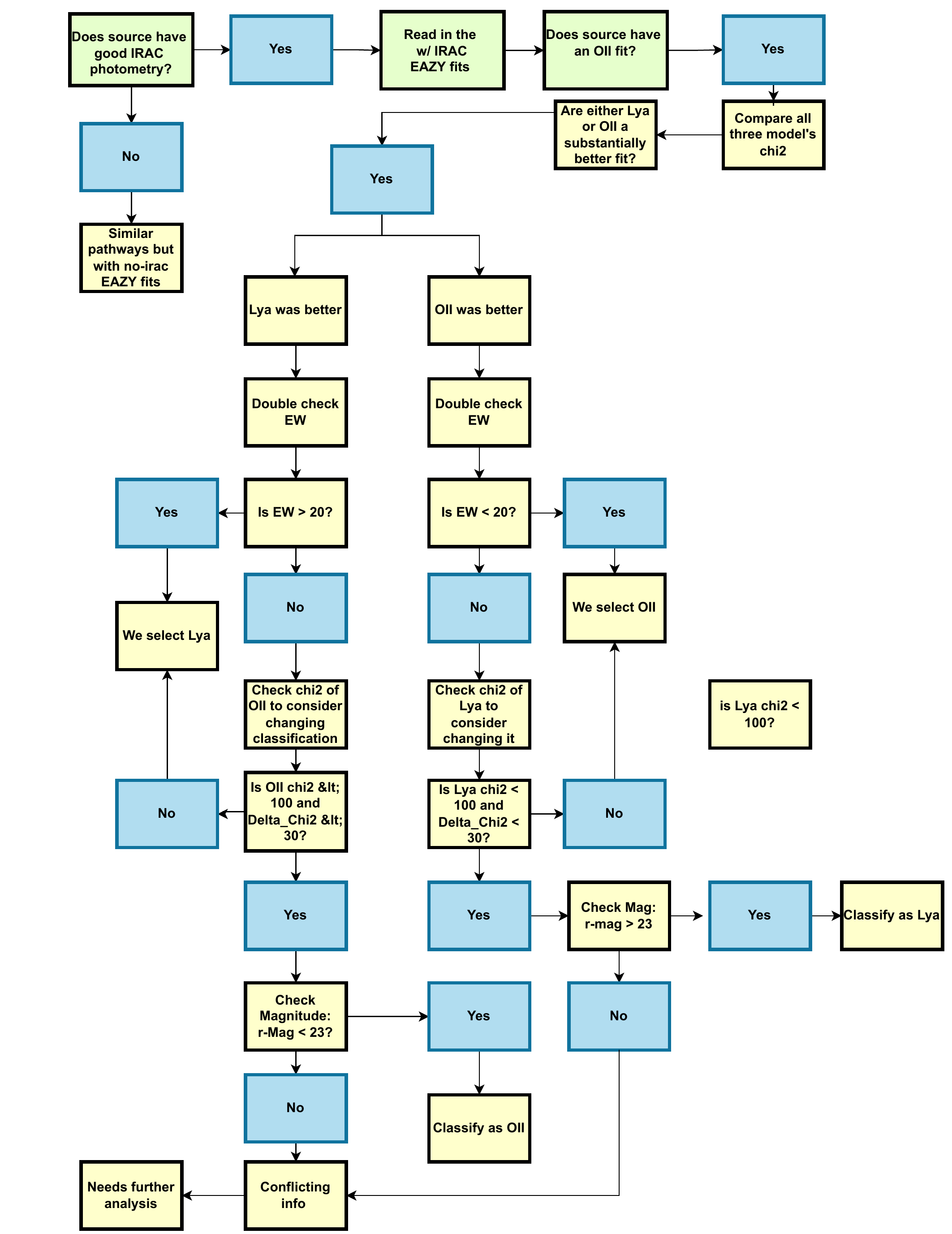}
    \caption{A flowchart diagram of the decision tree developed to classify our emission lines. It outlines the logic and routes sources take to select a line classification using photometric and spectroscopic information. The diagram shown here is the logical flow that a vast majority (90 \%) of the sources in this pilot study went through to get their emission line classified.}
    \label{fig:flow_chart}
\end{figure*}

\end{document}